\documentclass[preprint,12pt]{elsarticle}
\usepackage{graphicx}
\usepackage{caption}
\usepackage{subcaption}

\usepackage{booktabs,tabularx}
\usepackage[table]{xcolor}

\usepackage{amssymb}
\usepackage{amsmath}
\journal{Physica A: Statistical Mechanics and its Applications }

\begin{document}

\begin{frontmatter}

\title{Predicting pedestrian trajectories at different densities: A multi-criteria empirical analysis}

\author[label1]{Raphael Korbmacher}
\author[label2]{Huu-Tu Dang}
\author[label1]{and Antoine Tordeux}
\affiliation[label1]{organization={University of Wuppertal, Department for traffic safety and reliability},%Department and Organization
            city={Wuppertal},
            postcode={42119}, 
            country={Germany}}
\affiliation[label2]{organization={UMR 5505 IRIT, University Toulouse 1 Capitole},%Department and Organization
            city={Toulouse},
            %postcode={42119}, 
            country={France}}  
\begin{abstract}
Predicting human trajectories is a challenging task due to the complexity of pedestrian behavior, which is influenced by external factors such as the scene's topology and interactions with other pedestrians. A special challenge arises from the dependence of the behaviour on the density of the scene. In the literature, deep learning algorithms show the best performance in predicting pedestrian trajectories, but so far just for situations with low densities. In this study, we aim to investigate the suitability of these algorithms for high-density scenarios by evaluating them on different error metrics and comparing their accuracy to that of knowledge-based models that have been used since long time in the literature. 
The findings indicate that deep learning algorithms provide improved trajectory prediction accuracy in the distance metrics for all tested densities. Nevertheless, we observe a significant number of collisions in the predictions, especially in high-density scenarios. This issue arises partly due to the absence of a collision avoidance mechanism within the algorithms and partly because the distance-based collision metric is inadequate for dense situations.
To address these limitations, we propose the introduction of a novel continuous collision metric based on pedestrians' time-to-collision. Subsequently, we outline how this metric can be utilized to enhance the training of the algorithms.

%. which helps to asses how realistic the predictions are in terms of collision avoidance. Furthermore, we show how this error metrics can help to improve the training of the algorithm.
%The first metric is distance-based, while the second and third ones counts the number of collisions between pedestrians. Our findings reveal that deep learning algorithms provide improved trajectory accuracy in the distance metric, but knowledge-based models perform better in avoiding collisions. For future works we see a hybrid approach as useful, for with we present a continous error metric based on time-to-collison.
%We first observe, that the algorihtms outperform the modell in terms of accuracy in the distance metrics, but in the next step it is shown, that the predictions collide to often. Number of collisions is too high. THe state-of-the-art collision metrics has too many false-positives and is binary, which is why in the last step we present a new collision metric, that overcomes these flaws and can be used for the training of the algorihtms in future works.
\end{abstract}


\begin{highlights}
\item Empirical comparison of traditional physics-based models and state-of-the-art deep learning algorithms for predicting pedestrian trajectories.
\item The accuracy is tested across a range of different pedestrian densities.
\item Comparing by using three accuracy criteria: a Euclidean distance-based metric, a distance-based collision metric, and our novel time-to-collision based metric.
\item Improve the predictions by incorparating the time-to-collision based metric into the loss function of the deep learning algorithms.
%\item The paper conducts a robust, empirical comparison of traditional physics-based models and state-of-the-art deep learning algorithms for predicting pedestrian trajectories. 
%\item These methods are applied to a diverse range of data sets, which contain different pedestrian scenarios with different densities. 
%\item To obtain a thorough understanding of the relative strengths and weaknesses of each method, three comparison criteria are selected: a Euclidean distance-based metric, a distance-based collision metric, and our time-to-collision based metric.
%\item The comparison involving the collision metric reveals that deep learning algorithms underperform in the terms of collision avoidance. As a remedy, in the final phase, we incorporate the time-to-collision based metric into the loss function, resulting in a significant enhancement in prediction accuracy.
\end{highlights}

\begin{keyword}
Pedestrian trajectory prediction, deep learning, high-density, collision, time-to-collision



\end{keyword}

\end{frontmatter}


\section{Introduction}
\label{introduction}

The task of predicting pedestrian trajectories has emerged as a critical component in a variety of real-world applications, ranging from autonomous vehicles \cite{poibrenski2020m2p3} and human-robot interactions \cite{scheggi2014cooperative} to the design of events, infrastructure, and buildings \cite{bitgood2006analysis}, especially in the case of evacuation \cite{boltes2018empirical}. This topic has been addressed within academia from the two distinct disciplinary perspectives of pedestrian dynamics and data science \cite{korbmacher2022review}.

On one hand, the discipline of pedestrian dynamics applies a \emph{knowledge-based} (KB) approach, developing mathematical models that encapsulate the inherent rules governing pedestrian behavior \cite{chraibi2010generalized}. These models are utilized to conduct simulations in which pedestrian trajectories are computed. The difficulty lies in identifying fundamental mechanisms and parameters that induce realistic pedestrian behavior. On the other hand, computer scientists employ a \emph{data-based} (DB) approach, collecting extensive data and training sophisticated algorithms intended to predict pedestrian trajectories. Their focus predominantly lies in devising efficient algorithm architectures and meticulously fine-tuning the hyperparameters of DB algorithms.

While KB models cater to a broad array of applications and include macroscopic, mesoscopic, and microscopic models \cite{schadschneider2018pedestrian}, DB algorithms are primarily deployed for microscopic trajectory predictions in low-density scenes \cite{korbmacher2022review}. Low-density scenes denote situations with a medium pedestrian presence, where individuals possess a high degree of freedom and exhibit long-range interactions.
This paper seeks to explore the efficacy of DB algorithms in high-density situations and compare the results with those derived from traditional KB models. %, which have been conventionally used in these scenarios. 
One significant challenge inherent in such a comparison lies in devising a fair and comprehensive evaluation. While prior studies have demonstrated the superior performance of DB algorithms in terms of prediction accuracy, these evaluations have exclusively pertained to low-density data \cite{alahi2016social}, focusing solely on distance metrics such as Average Displacement Error (ADE) \cite{pellegrini2009you} and Final Displacement Error (FDE) \cite{lerner2007crowds}. We propose to extend this evaluation by incorporating two additional metrics: a binary distance-based collision metric as proposed by Kothari et al. \cite{kothari2021human}, and an original continuous time-to-collision-based metric. % that we developed.\\

Findings indicate that the DB algorithms surpass the KB models across all tested densities in terms of distance metrics. However, the DB algorithm predictions generate a significantly higher number of collisions when compared to the real trajectories and the KB models, which are typically designed with collision avoidance mechanisms.

\section{Related work}
\label{rel work}
In the following chapter, an extensive review of the KB models Section~\ref{KB_models} and DB algorithms Section~\ref{DB_alg} is conducted to provide a robust foundation for the current study.

\subsection{Knowledge-based models}
\label{KB_models}
KB models  have a rich history in pedestrian dynamics that dates back to the middle of the 20th century. These models apply principles from physics, such as force fields and particles, to understand and predict the behavior of pedestrians. Currently, KB pedestrian models range from macroscopic, mesoscopic, and microscopic models among others modeling scale characteristics. Macroscopic and mesoscopic approaches are borrowed from continuous fluid dynamics or gas-kinetic models describing the dynamics at an aggregated level, while microscopic approaches model individual pedestrian motions \cite{hoogendoorn2014continuum}. For pedestrian trajectory predictions macroscopic and mesoscopic models are less relevant, which is why in the following we focus on the microscopic models.

In these models the individual pedestrian behavior is
described according to certain %KB 
rules and mechanisms ground on physical social, or psychological factors \cite{chraibi2018modelling}. These rules and mechanisms are formulated in hand-crafted dynamic equations based on Newton’s laws of motion. Given the input information about the initial status of the pedestrians like position, velocity, and acceleration a forward simulation of KB models %these rules 
can be used to predict the future trajectories. Depending on the modelling order %inputs and outputs 
of the model, they can be classified into \emph{decision-based} (zeroth order), \emph{velocity-based} (first order), and \emph{acceleration-based}  models (second order) \cite{korbmacher2022review}.
In acceleration-based models, typically force-based models, the movement of pedestrians is defined by a superposition of exterior forces. Most acceleration-based models are force-based consisting of a relaxation %(or anisotropic) 
term to the desired direction and an interaction term \cite{korbmacher2022review}. This last term is generally the sum of repulsion with the neighbours and obstacles. This is also the case in the most famous force-based model, the \emph{Social Force model} (SF) from Helbing and Molnar \cite{helbing1995social}, where the interaction force is an exponential gradient of a distance-based potential. % on the distances and the speed differences with the neigbors.\\

Velocity-based models are speed functions, depending
on the position differences with neighbors and obstacles. In opposite to the acceleration-based models that are based on second-order differential equations, the velocity-based models rely on first-order equations. Many of these models are based on collision avoidance techniques and are formulated as optimisation problems on some ensemble of feasible trajectories devoid of collisions. The most famous models are the Reciprocal Velocity Obstacle model \cite{van2008reciprocal,xu2019generalized} and the \emph{Optimal Reciprocal Collision Avoidance} (ORCA) \cite{van2011reciprocal}.

In the last class of models, the decision-based or rule-based models, the pedestrian behavior is not modeled based on differential equations, but on rules or decisions determining the new agent positions, velocities, etc \cite{chraibi2018modelling}. The time is considered to be discrete for this class of decision-based models, which are typically Cellular automata \cite{burstedde2001simulation,lovreglio2015calibrating,blue1999cellular,kirchner2002simulation}.

KB models in pedestrian dynamics encompass various approaches, including microscopic models, which can be used for trajectory predictions. Most of these models focus on the interactions between pedestrians and the environment and, as a result, are fundamentally based on collision avoidance mechanisms.

\subsection{Data-based algorithms}
\label{DB_alg}
The previously outlined approach is fundamentally grounded in a theoretical modeling framework. Essential mechanisms are pre-identified and expressed in the form of equations, equipped with a handful of significant parameters that require calibration and validation. The operational functionality of these models extends to the simulation of pedestrian scenes, enabling the prediction of future trajectories as a consequential byproduct. In the DB approach, the prediction of trajectories is not a secondary outcome, but the main objective. 
The parameters (or coefficients) of the algorithm have no physical meaning and can not be interpreted.  
These parameters are determined through a process of training the algorithm with data, with the goal of minimizing a predefined cost function. The common cost function for trajectory predictions is the displacement error %distance-based 
metrics ADE or FDE. %After training the performance of the algorithm is examined with testing data, which was previously not used to train the algorithm.
The trained algorithms are then tested on new data, i.e. data which was previously not used to train the algorithm (cross-validation). 

Over the past decade, a multitude of studies employing various data-based methodologies have been published with the aim of predicting pedestrian trajectories. For a comprehensive overview, please refer to the following reviews %references 
\cite{korbmacher2022review,kothari2021human,rudenko2020human, bighashdel2019survey}.
The majority of these studies use supervised deep learning techniques with either \emph{Long Short-Term Memory} (LSTM) or \emph{Generative Adversarial Networks} (GAN) architectures. Among the most influential works in this field are \emph{Social-LSTM} by Alahi et al. \cite{alahi2016social} and \emph{Social-GAN} by Gupta et al. \cite{gupta2018social}. These groundbreaking papers have served as the inspiration for a number of subsequent studies, which have extended these initial algorithms to incorporate elements such as scene information \cite{xue2018ss}, attention mechanisms \cite{haddad2019situation}, graph neural networks \cite{ huang2019stgat, monti2021dag}, and heterogeneity among pedestrians \cite{lai2020trajectory}.
In addition, it's noteworthy to mention the utilization of convolutional neural networks, which can be trained on video data rather than trajectory data \cite{yi2016pedestrian, nikhil2018convolutional,chen2020pedestrian}. There are also studies using reinforcement learning algorithms, which do not require explicit data, but rather operate based on a defined reward function \cite{chen2017socially,wan2018robot}.

%https://www.mdpi.com/1424-8220/19/5/1223

%2 artikel von Kothari

\section{Method}
\label{Methodology}

In this section, we provide a detailed description of the datasets used in this study (Section \ref{data_set}). We then define the models and algorithms employed in the following analysis (Section \ref{Models&Algorithms}) and introduce distance and collision-based evaluation metrics that will be used to assess the performance of the different models and algorithms (Section~\ref{Evaluation}).

\subsection{Pedestrian trajectory data}
\label{data_set}
In recent years, a large number of datasets have been %generated 
collected and made publicly available from an extensive range of studies, which mainly include real-world trajectories of scenarios with low pedestrian densities ranging from 0.1 to 0.4 ped/m$^2$. For a comprehensive overview, one may refer to \cite{amirian2020opentraj}. Interestingly, datasets corresponding to higher-density situations are noticeably absent, possibly due to the challenges associated with the data collection (i.e. trajectory extraction). %their automatic tracking. 

\subsubsection{Low-density dataset}
Initially, we assess the performance of the models and algorithms using low-density datasets, which typically feature long-range interactions and scenarios involving less than 0.5 ped/m$^2$ \cite{korbmacher2022review}. Pedestrians in these scenes have high degrees of freedom, and their behaviour is primarily influenced by a few neighbouring people.

\begin{figure}[htbp]
    \centering
    \begin{subfigure}[t]{0.4\textwidth}
        \centering
        \includegraphics[width=2.3in]{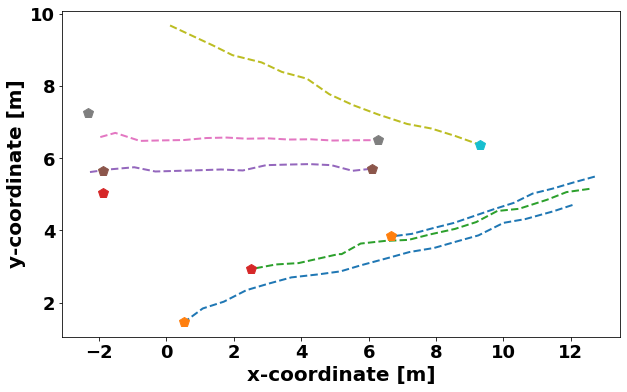}
        \caption[1]{Example from ETH.}
    \end{subfigure}%
    ~~~~~~~ 
    \begin{subfigure}[t]{0.4\textwidth}
        \centering
        \includegraphics[width=2.3in]{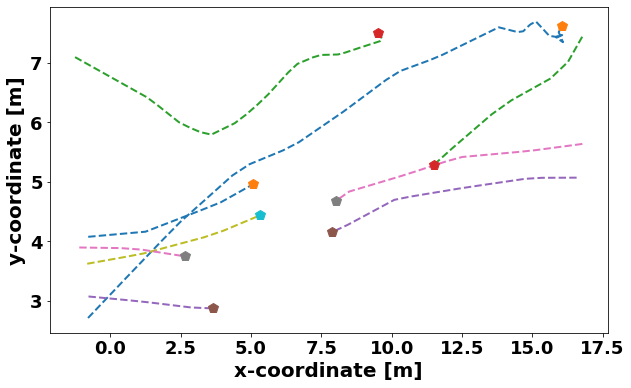}
        \caption{Example from ZARA02.}
    \end{subfigure}
    \caption{Illustrative examples of trajectory samples from low-density datasets (ETH and UCY sets).}
    \label{Example_low}
\end{figure}

Given their emergence as benchmark datasets in pedestrian studies over recent years, we have selected the ETH \cite{pellegrini2009you} and UCY \cite{lerner2007crowds} datasets for the analysis. The ETH dataset comprises a total of 750 trajectories, divided into two subsets: ETH and Hotel. Fig.~\ref{Example_low} (a) shows an example of a segment from the hotel dataset. The UCY dataset, on the other hand, has been subdivided into three subsets: ZARA01, ZARA02, and UNIV, collectively containing 786 trajectories. An example of ZARA02 is shown in Fig.~\ref{Example_low} (b). Both datasets, collected in outdoor environments, encapsulate a variety of pedestrian traffic patterns, including unidirectional, bidirectional, and multidirectional. These datasets have been recorded at a framerate of 2.5 frames per second.

\subsubsection{High-density datasets}
High-density data refers to pedestrian situations characterized by more than 2 ped/m$^2$, commonly known as crowds. The generation of accurate pedestrian trajectory data in these situations is a challenging task due to the difficulties in the automatic pedestrian identification and tracking. % correctly detecting single pedestrians and following them through time. 
The problems that usually occur are that the pedestrians often gather together, occlude each other, and result in overlapping in pedestrian shapes \cite{wang2018repulsion}. 
These factors contribute to the scarcity of real-world high-density pedestrian trajectory datasets. Nevertheless, there are some rare examples available, which can be found in \cite{johansson2009analysis,polus1983pedestrian,mori1987new}. In addition to these real-world datasets, Forschungszentrum Jülich has conducted various laboratory experiments, including HERMES, BaSiGo, CroMa, and CrowdDNA, which provide high-density trajectory data \cite{ExpJuelich, cao2017fundamental, seyfried2005fundamental}. Furthermore, other experimental datasets can be found in \cite{echeverria2021estimating, zanlungo2023macroscopic}.

\begin{figure}[htbp]
    \centering
    \begin{subfigure}[t]{0.45\textwidth}
        \centering
        \includegraphics[width=2.27in]{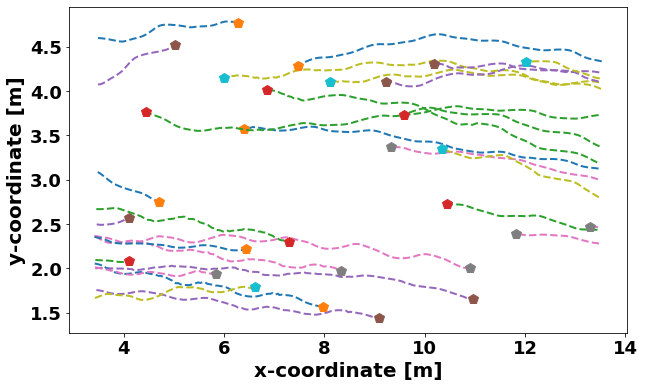}
        \caption[1]{Example from bidi3.}
    \end{subfigure}%
    ~ 
    \begin{subfigure}[t]{0.45\textwidth}
        \centering
        \includegraphics[width=2.3in]{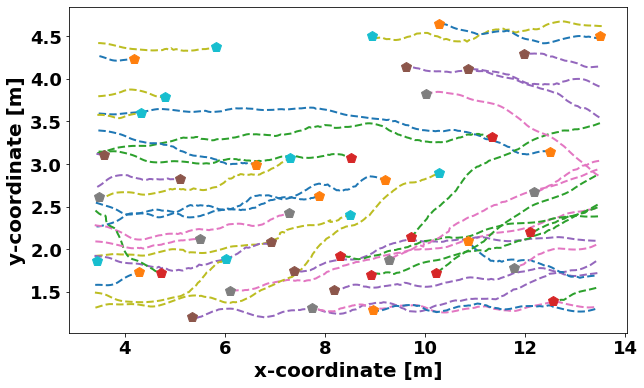}
        \caption{Example from bidi5.}
    \end{subfigure}
    \caption{Illustrative examples of trajectory samples from high-density datasets (Juelich experiments).}
    \label{Example high}
\end{figure}

In this study, we primarily focus on the corridor experiments with bidirectional flow from the Forschungszentrum Jülich, as they encompass a diverse array of interactions. 
The experiment setup incorporates two starting points or entrances, from where pedestrian start their walk. The size of the recording area is a = 10 m, and $ b_{corr} $ = 4 m. We utilize data gleaned from six distinct bidirectional corridor settings, as showcased in Table~\ref{average_density}. For illustrative purposes, we depict the trajectories from the experiment exhibiting the third-highest density (bidi3), alongside those from the fifth-highest density (bidi5) in Fig.~\ref{Example high}. In total, the dataset comprises 3096 trajectories, recorded at a framerate of 16 frames per second.

% For more customizable captions

\begin{table}[ht]
\centering
\rowcolors{4}{lightgray!25}{white}
\caption{Overview of pedestrian trajectory datasets.}
\captionsetup{skip=0.5\baselineskip}
\begin{tabular}{cccccc}     
\toprule
\textbf{Dataset} & \textbf{Setting} & \textbf{Number of} & \textbf{Average} & \textbf{Maximum} \\
& & \textbf{Pedestrians} & \textbf{Density} & \textbf{Density} \\
\hline\noalign{\smallskip}\hline\noalign{\smallskip}
\hspace{-1cm}\bf{ETH} & & & & \\
\hspace{0.5cm}\quad ETH & Outdoor &  365 & 0.14 & 0.35 \\
\hspace{0.5cm}\quad HOTEL & Outdoor  & 420 & 0.13 & 0.32 \\ 
\midrule
\hspace{-1cm}\bf{UCY} & & & & \\
\hspace{0.5cm}\quad ZARA01 & Outdoor & 148 & 0.21 & 0.51 \\
\hspace{0.5cm}\quad ZARA02 & Outdoor & 204 & 0.27 & 0.48 \\
\hspace{0.5cm}\quad UNIV & Outdoor  & 434 & 0.38 & 0.52 \\ 
\midrule
\hspace{-0.5cm}\bf{JUELICH} & & & & \\ 
\hspace{0.5cm}\quad bidi1 & Lab &  141 & 0.38 & 0.55 \\
\hspace{0.5cm}\quad bidi2 & Lab &  259 & 0.58 & 0.75 \\
\hspace{0.5cm}\quad bidi3 & Lab &  480 & 1.00 & 1.15 \\
\hspace{0.5cm}\quad bidi4 & Lab &  743 & 2.32 & 3.03 \\
\hspace{0.5cm}\quad bidi5 & Lab &  643 & 2.64 & 3.275 \\ 
\hspace{0.5cm}\quad bidi6 & Lab &  830 & 3.0 & 3.775 \\ 
\bottomrule
\end{tabular}

\label{average_density}
\end{table}

\subsection{Models and Algorithms}
\label{Models&Algorithms}
In the subsequent trajectory predictions, we focus on two crucial elements: predictions across varying pedestrian densities and the utilization of diverse models/algorithms. We systematically use the notation $x_i\in\mathbb R ^2$ and $v_i\in\mathbb R^2$ to represent the position and velocity of the $i$-th pedestrian. The Euclidean distance is denoted as $|\cdot|$, while $x$ and $v$ refer to the vectors of pedestrian positions and velocities. These vectors have a dimension of $2N$ for $N$ pedestrians. All variables, including $x(t)$ and $x_i(t)$, depend on the time $t$.

We first select two contemporary knowledge-based models to facilitate this. The first is the Social Force model (SF) by Helbing and Molnar \cite{helbing1995social}, and the second is the Optimal Reciprocal Collision Avoidance (ORCA) approach by van den Berg et al.\ \cite{berg2011reciprocal}.
The SF model is a widely adopted method that simulates pedestrian movement, treating individuals as particles influenced by various forces. The acceleration within this model is calculated from the summation of three forces as demonstrated in Equation \ref{social_force_equ} %These forces include personal motivations, such as reaching a destination, and social interactions, like avoiding collisions with other pedestrians or objects. The model takes into account the balance between the desire to reach a goal and the need to maintain a comfortable distance from others.
% \begin{equation}
%     \frac{d}{dt} v_i &= D_i(v_i) + I_i(t,x,v) + B(t,x_i,v_i)
% \label{social_force_equ}
% \end{equation}
% Here, $D_i$ denotes the driving force that pedestrian i experiences to achieve their desired speed and direction. $I_i$ represents the summation of the social forces derived from the repulsive effects of pedestrians maintaining distance from each other. The third force, $B$, denotes the cumulative interaction forces between pedestrian i and obstacles.\\

\begin{equation}
\label{social_force_equ}
m_i \frac{dv_i}{dt} = m_i \frac{v^0_i%e^0_i
- v_i}{\tau} + \sum_{j \neq i} \nabla U(x_j-x_i) + \sum_{W} \nabla V(x_W-x_i)
\end{equation}
where $m_i$, $v_i$ and $v^0_i$%, and $e^0_i$ 
are the mass, current velocity and preferred velocity %speed, desired direction 
of the $i$-th pedestrian, respectively, while $U$ and $V$ are distance-based interaction potential, e.g.,
\begin{equation}
    U(d)=Ae^{-\|d\|/B},\qquad A,B>0.
\end{equation}

The first term of Equation \ref{social_force_equ} denotes the driving force that the $i$-th pedestrian experiences to achieve their desired speed and direction within the reaction time $\tau >0$. The second term represents the summation of the social forces derived from the repulsive effects of pedestrians maintaining distance from each other. The third force denotes the cumulative interaction forces between pedestrian i and obstacles.

In contrast, ORCA is centered on efficiently determining collision-free velocities for multiple agents within a shared environment  \cite{berg2011reciprocal}. It is based on a geometric approach to model agent interaction, identifying the range of velocities that guarantees collision avoidance within a specified time horizon by extrapolating linearly (i.e., assuming constant the velocity) the trajectories. These computations result in collision cones between the pedestrians, that can be empty. 
Further mechanisms are taken into account to avoid unrealistic oscillation effects and makes the agents acting independently without communicating with each other.
Ultimately, each agent subsequently selects the optimal velocity $v_i$ that is closest to its ideal %corrected 
(preferred) velocity $v^0_i$ within the feasible velocity region \cite{narang2015generating} excluding collisions, as shown in Equation~\ref{eq:VO1}
\begin{align}
\label{eq:VO1}
    v_i(t+dt)=\text{arg}\hspace{-7mm}\min_{v\,\in\, \cap_{j\ne i}\text{ORCA}_{ij}(t)}\|v-v^0_i\|,
\end{align}
with $\text{ORCA}_{ij}$ the set of feasible (collision-free) velocities of the $i$-th pedestrian with the $j$-th neighboring pedestrian, and $dt$ a (small) time step (typically equal to $dt=0.01$\,s). 
% Here, the collision cones depend on the current velocities of the agents. 
% This relationship makes the velocity model implicit. 
% In practice, the implicit system is solved numerically using semi-implicit numerical schemes combining an explicit Euler solver for the velocity 
% \begin{align*}
%    v_i(t+dt)=\text{arg}\hspace{-5.5mm}\min_{v\,\not\in\, \cup_{j\ne i}\text{VO}_{ij}(t)}\|v-u_i(t)\|^2,&&i=1,\ldots,n,
% \end{align*}
% with an implicit Euler scheme for the positions of the agents \cite{berg2011reciprocal,van2008reciprocal}.

% \subsubsection{Deep learning algorithms}
For the DL approach, we adopt the LSTM network, which is widely used in pedestrian trajectory prediction. In this architecture, we leverage two algorithms. The first, referred to as the Vanilla LSTM, considers the historical trajectory over $[t-T_o,t]$, with $T_o>0$ the observation time, to predict the trajectory over $[t,t+T_p]$ with $T_p$ the prediction time
\begin{equation}
x_i(t+t_p) =\text{LSTM}\big(t+t_p,\big(x_i(t-t_o),~t_o \in \mathbb [0,T_o]\big)\big),\qquad \forall t_p \in \mathbb [0,T_p] .
\label{vanilla_lstm_eq}
\end{equation}

This algorithms is grid-based which means that the input is discretised in a local grid constructed around the pedestrian. The second algorithm, known as the Social-LSTM \cite{alahi2016social}, incorporates a social pooling mechanism to aggregate information about neighboring entities within the grid, as illustrated in Equation \eqref{social_lstm_eq}
\begin{equation}
x_i(t+t_p) =\text{SLSTM}\big(i,t+t_p,\big(x(t-t_o), ~t_o \in \mathbb [0,T_o]\big)\big), \qquad \forall t_p \in \mathbb [0,T_p].
\label{social_lstm_eq}
\end{equation}
With this mechanism, the model can use the historical trajectories of surrounding pedestrians $x$ over $[t-T_o,t]$, enabling the consideration of interactions in the predictions. Table~\ref{tab:models_and_algorithms} lists the various types of information required by each model or algorithm to make predictions.
The first approach in the Table~\ref{tab:models_and_algorithms} is the constant velocity model. It is the most simple approach making prediction assuming the pedestrian velocities remain constant

\begin{equation}
x_i(t + t_p) = x_i(t) + t_p v_i(t), \qquad \forall t_p \in \mathbb [0,T_p].
\end{equation}

\subsection{Implementation details}
The DL algorithms are trained with a learning rate of 0.0015, and a RMS-prop is used as the ADAM optimizer. The batch size is 8, and we train for 12 epochs. As a loss function, the mean squared error is used. For the validation and testing, we use a hold-out validation strategy. 15 \% of the data is used for validation, 15 \% for testing and the rest for training. The computations are performed using the PyTorch library\footnote{http://pytorch.org}. Two different observation and prediction times are employed in the study. For 1.2-second predictions, a 1-second observation period is utilized. For 4.8-second predictions, the observation length extends to 3.6 seconds.
We employ the Stochastic Gradient Descent (SGD) algorithm to fit the parameters of the KB models to the training data by minimizing the ADE metric. For the SF, we optimize the preferred velocity, the interaction potential, and the reaction time, according to \cite{kreiss2021deep}. For the ORCA we optimize the distance to pedestrians that are taken into account and the corresponding reaction time. It is worth noting that the reaction time plays a significant role in the model's behavior. A shorter reaction time prompts a quicker response to the presence of other agents but reduces the pedestrian's freedom in choosing their velocities, as mentioned in \cite{van2011reciprocal}. On the other hand, the CV model stands apart as it does not require any calibration or training due to its parameter-free nature. An overview of the most strinking differences between the approaches is presented in Table~\ref{tab:models_and_algorithms}. The displayed information includes the input of primary pedestrians and their neighbors, as well as the optimization method and the parameters being optimized.

\begin{table}[!ht]
\caption{Overview of important features.}
\label{tab:models_and_algorithms}       
\rowcolors{2}{lightgray!25}{white}
\begin{tabular}{p{1.8cm}p{3cm}p{1.8cm}p{3.5cm}p{1.5cm}}
\hline\noalign{\smallskip}
\bf{Approach} & \bf{Primary Input} & \bf{Neighbor Input} & \bf{Optim. Parameters} & \bf{Optim.\ Method} \\
\hline\noalign{\smallskip}\hline\noalign{\smallskip}
CV & Current State & None & None & None\\
SF & Current State & Yes & Preferred Velocity, Interaction Potential, Reaction Time & SGD \\
ORCA & Current State & Yes & Neighbor Distance, Reaction Time & SGD \\
Vanilla-LSTM & Past Trajectory & None & Large number of coefficients & ADAM\\
Social-LSTM & Past Trajectory & Yes & Large number of coefficients & ADAM\\
\noalign{\smallskip}\hline\noalign{\smallskip}
\end{tabular}
\end{table}

\subsection{Evaluation}
\label{Evaluation}

\subsubsection{Distance-based metrics}
A crucial question that emerges when employing pedestrian trajectory prediction algorithms in high-density scenarios is the appropriate method of evaluation. This query is applicable more broadly to the field of pedestrian dynamics. Without an objective metric for evaluation, it is impossible to definitively determine the best-fitting model or algorithm.
In low-density trajectory predictions, two metrics based on Euclidean distance are commonly utilized. The first is the Average Displacement Error (ADE) \cite{pellegrini2009you}, which measures the distance between the predicted trajectory and the ground truth trajectory at a set number of points
\begin{equation}
    \textbf{ADE}=\frac{1}{N T}\sum_{i=1}^{N} \sum_{t=1}^{T}\|\hat x_i(t)-x_i(t)\|,
\label{m1}
\end{equation}

$x_i(t)$ being the actual position of the $i$-th pedestrian at time $t$ while $\hat x_i(t)$ is the predicted position.

\subsubsection{Discrete distance-based collision metric}
These two distance metrics are effective in guiding the algorithm to make predictions that match actual trajectories. However, a significant drawback is their focus on distance, leading to an underestimation of the repulsive forces that arise between pedestrians. Consequently, these metrics do not account for potential overlaps or collisions between pedestrians. 

To tackle this problem, the following distance-based collision metric has been presented by Kothari et al.\ \cite{kothari2021human}
\begin{equation}
	\textbf{Col} = \frac{1}{|S|}\displaystyle \sum_{\hat{Y} \in S} Col(\hat{Y}),
\label{m3}
\end{equation}
with 
\begin{equation}
    Col(\hat{Y})=\min\bigg(1,\sum_{t=1}^{T}\sum_{i=1}^{N}\sum_{j>i}^{N} \big[||\hat{x}_i(t)-\hat{x}_j(t)|| \leq 2R\big]\bigg),
    \label{m3b}
\end{equation}
where $S$ includes all scenes in the test set, $\hat{Y}$ represents a scene prediction containing $N$ agents, and $\hat{y}_i$ is the prediction of agent $i$ over the prediction time of $T$, while $[\cdot]$ is the Iverson bracket
\begin{equation}
    \big[P\big]=\left\{\begin{array}{ll}
    1&\text{if $P$ is true,}\\
    0&\text{otherwise.}
    \end{array}\right.
\end{equation}
This last metric counts a prediction as a collision when a predicted pedestrian trajectory intersects with neighboring trajectories, thus indicating the proportion of predictions where collisions occur. A vital factor in this calculation is the chosen pedestrian size (radius), represented by the variable $R$ in equation~\eqref{m3b}. An increase in $R$ will likewise increase the number of collisions. 

\subsubsection{New TTC-based error metric}
\label{ttc_metric}
The collision metric Col discussed so far has been designed to mitigate overlapping and collisions between pedestrians. This is based on the principle that pedestrians inherently strive to avoid physical contact with others, an aspect not sufficiently captured by ADE. Nevertheless, this collision metric isn't without its shortcomings, which we aim to address. One drawback is that the metric is based on binary collision identifications. As such, it doesn't distinguish between minor instances of contact, such as a shoulder brush between two pedestrians walking side-by-side, and significant collisions such as a head-on crash. Another limitation is its inability to account for scenarios where a prediction results in multiple collisions. Additionally, the metric models pedestrians as circles, represented by radius $R$, although an elliptical representation would be more accurate.
The proposed solution is to introduce a collision metric based on the concept of Time-to-Collision (TTC) between two pedestrians. In this system, a low TTC implies an impending collision.

The TTC is estimated as the time remaining before two moving pedestrians, donated as $i$ and $j$, collide based on their current velocities. Suppose that $R_i$ and $R_j$ are the radius of $i$-th and $j$-th pedestrians, respectively. Given the relative distance and relative velocity between the two pedestrians 
\begin{equation}
    x_{ij} = x_i - x_j\qquad\text{and}\qquad v_{ij} = v_i - v_j,
\end{equation}
a collision between $i$-th and $j$-th pedestrians occurs if there exists a time $\tau > 0$ such that $x_{ij} + v_{ij}\tau$ lies within a circle centered at $(0, 0)$ with radius $2R$. %$R_i+R_j$. 
Mathematically, this condition can be expressed as $ \|x_{ij} + v_{ij}  t \| < 2R$ where $\|\cdot\|$ denotes Euclidean norm. It turns out to solve the quadratic inequality in $t$, and $\tau$ is the smallest positive root:
\begin{equation}
	\tau_{ij} = \frac{\displaystyle -x_{ij} \cdot v_{ij} - \sqrt{(x_{ij} \cdot v_{ij})^2 - || v_{ij} ||^2 (||x_{ij}||^2 -  4R^2)} }{|| v_{ij} ||^2}.
\label{ttc}
\end{equation}

In this scheme, if no collision is imminent, then $\tau_{ij}$ is not real-value or is negative. We set in this case $\tau_{ij} = \infty$. 
Conversely, we assume by convention that $\tau_{ij}= 0$ if the pedestrians are already in collision.
To be able to compare the performances of different prediction approaches, the inverse of average TTC (ITTC) is calculation according to 
\begin{equation}
\begin{aligned}
	ITTC = \frac{N T}{\displaystyle\sum_{i=1}^{N} \sum_{t=1}^{T} \min_{j\ne i} \big\{\tau_{ij}(t),\tau_\text{max}\big\}},
% \sum_{j \in Nb} \min(ttc, ttc < \tau).
\label{col-I}
\end{aligned}
\end{equation}
with $\tau_\text{max}=12$\ seconds a maximal TTC threshold value.

\section{Results}
\label{Results_distance}

In the upcoming chapter, the objective is to showcase the effectiveness of different approaches in making predictions across diverse time intervals and densities. To accomplish this, we will conduct comprehensive evaluations of the predictions using ADE (Section~\ref{distance_metric}), the distance-based collision metric Col (Section~\ref{Distance-based collision metric}), and the ITTC (Section~\ref{TTC-based collision metric}). Additionally, in Section~\ref{TTC Metric for improving performance of the algorithm}, we will demonstrate how the TTC metric can be leveraged to enhance trajectory predictions.

\subsection{Distance metric}
\label{distance_metric}
\begin{figure}[htbp]
    \centering
    \begin{subfigure}[t]{0.4\textwidth}
        \centering
        \includegraphics[width=2.3in]{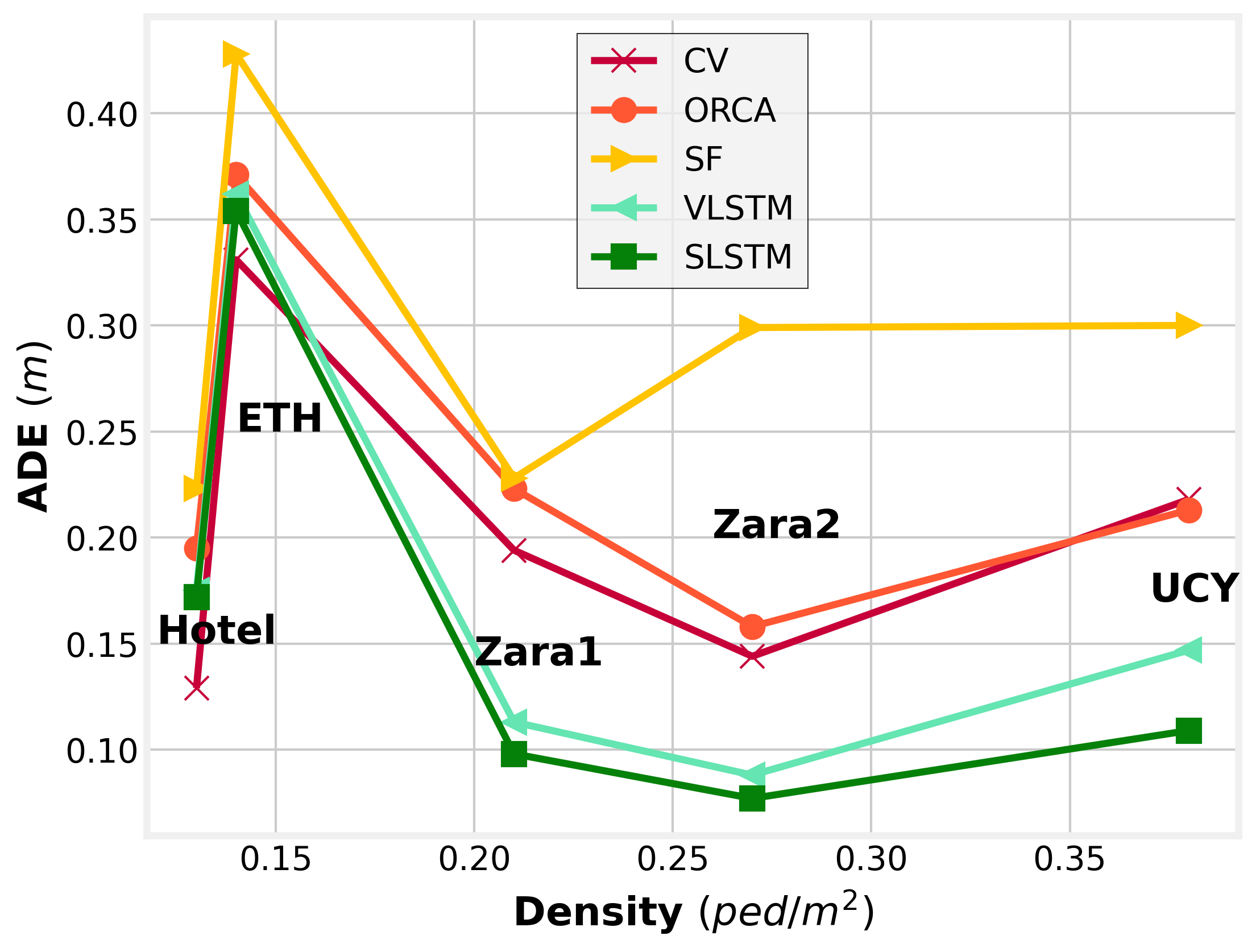}
        \caption[1]{Low-density data with $T_p=1.2$ sec.}
    \end{subfigure}%
    \hspace{1cm}
    \begin{subfigure}[t]{0.4\textwidth}
        \centering
        \includegraphics[width=2.3in]{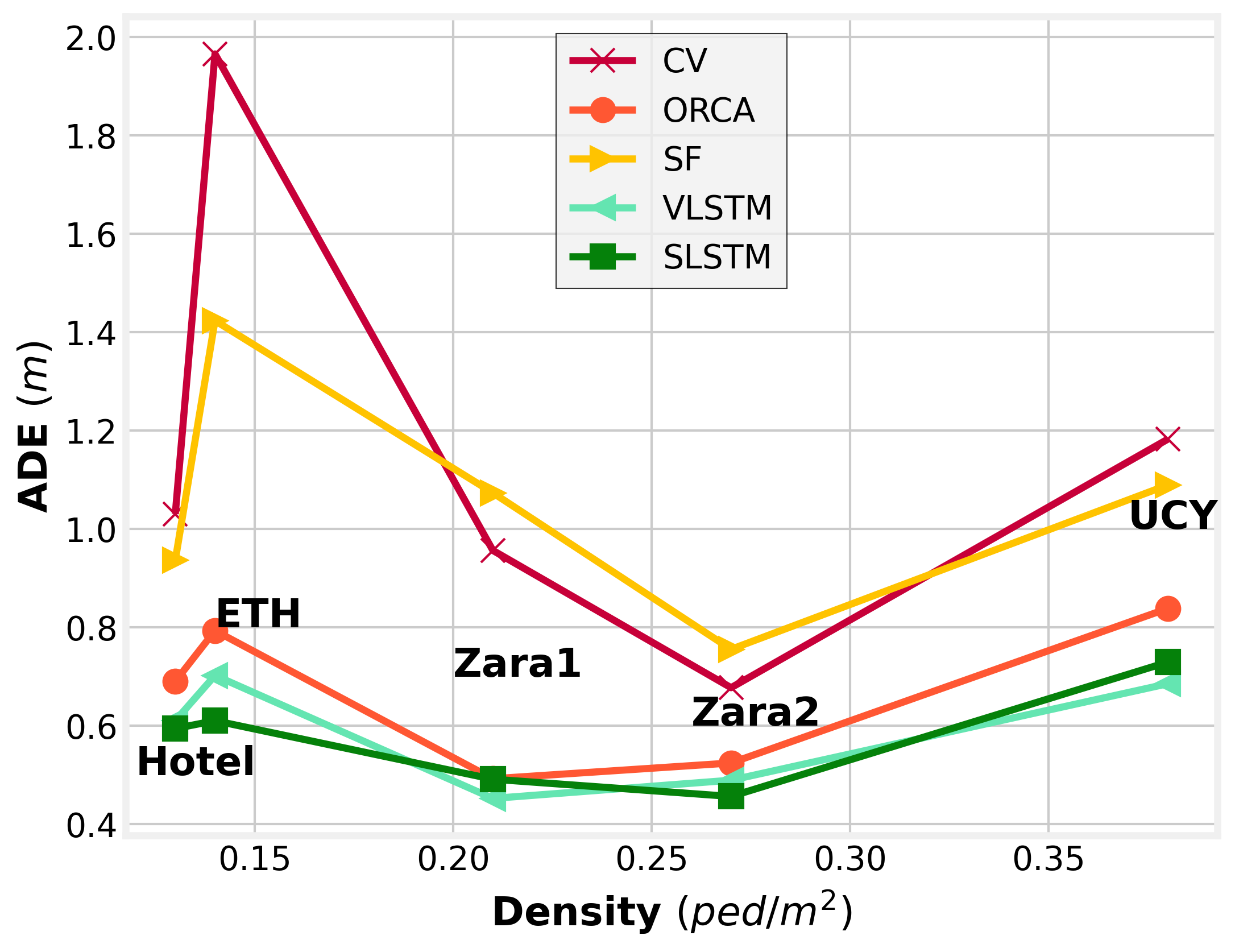}
        \caption{Low-density data with $T_p=4.8$ sec.}
    \end{subfigure}
    \caption{Distance error-metric (ADE) for low-density datasets.}
    \label{ADE_low}
\end{figure}

At first, we will focus on analyzing the distance-error metrics for the low-density datasets. The outcomes of the predictions are presented in Fig.~\ref{ADE_low}. As previously mentioned, we utilize five datasets for the low-density predictions, with densities ranging between 0.13 and 0.38 ped/m$^2$. In Fig.~\ref{ADE_low}, the x-axis displays the average densities of each dataset, while the y-axis represents the ADE metric. On the left side a prediction time $T_p$ of 1.2 seconds is chosen and on the right side a prediction time of 4.6 seconds.

The Fig.~\ref{ADE_low} initially demonstrates that the algorithms consistently outperform the models across nearly all low-density datasets, showcasing their superior predictive capabilities. The Social-LSTM exhibits slightly better performance compared to the Vanilla-LSTM.\\
Furthermore, notable differences can be observed between the two prediction horizons. While the CV approach performs reasonably well at shorter prediction times, the error increases significantly, nearly doubling compared to the other approaches, as the prediction time extends. The substantial error of the CV approach on the ETH dataset highlights its challenging nature, with the highest deviation from keeping speed and direction constant. Interestingly, the most complex approach, namely SLSTM, demonstrates the best performance on this dataset, indicating its effectiveness in handling complexity. Moreover, it is worth noting that an increase in density does not necessarily result in higher prediction errors for the low-density dataset. For instance, Zara02, despite having a higher density than ETH, exhibits a lower average ADE. 

\begin{figure}[htbp]
    \centering
    \begin{subfigure}[t]{0.4\textwidth}
        \centering
        \includegraphics[width=2.3in]{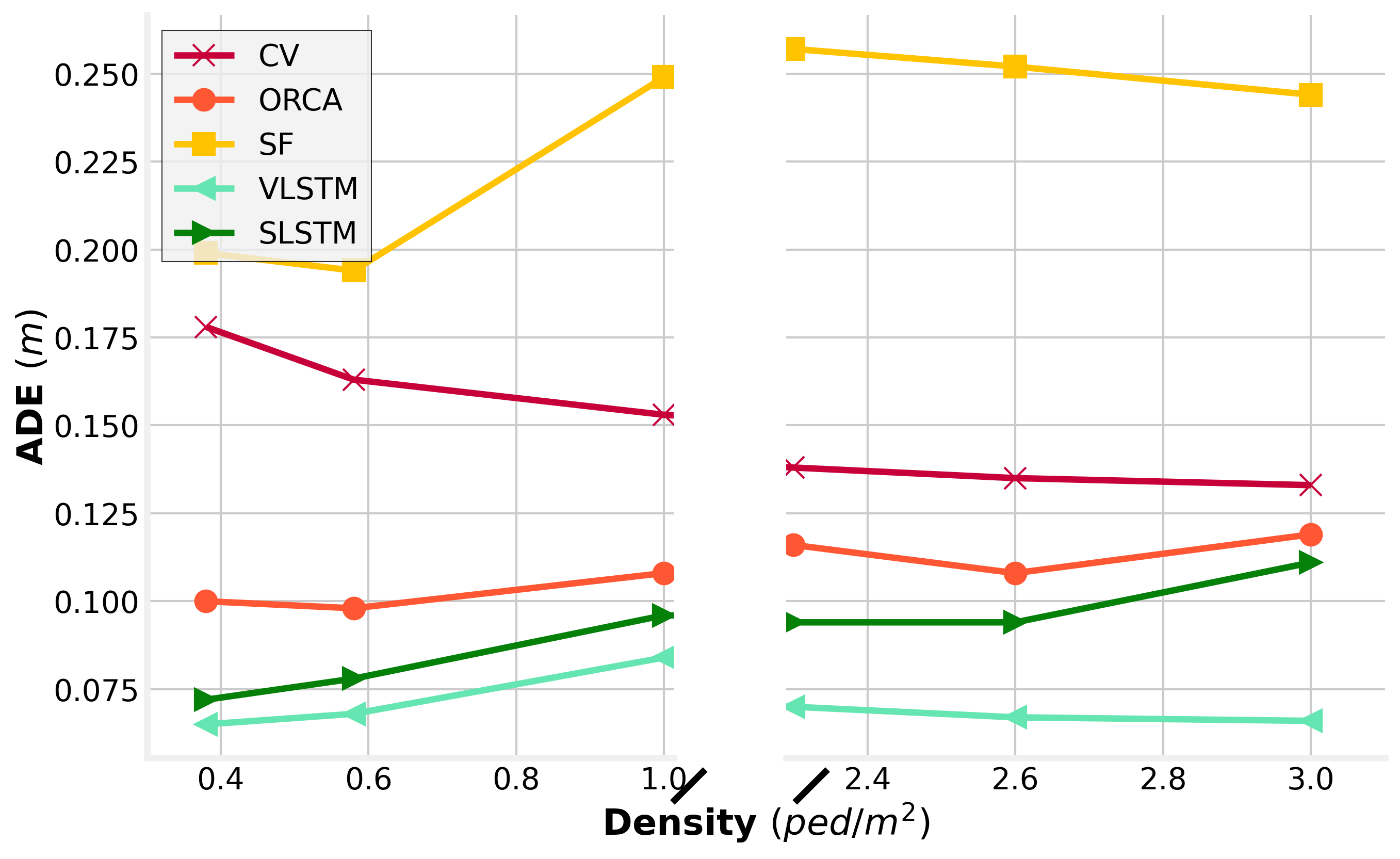}
        \caption[1]{High-density data with $T_p=1.2$ sec.}
    \end{subfigure}%
    \hspace{1cm}
    \begin{subfigure}[t]{0.4\textwidth}
        \centering
        \includegraphics[width=2.3in]{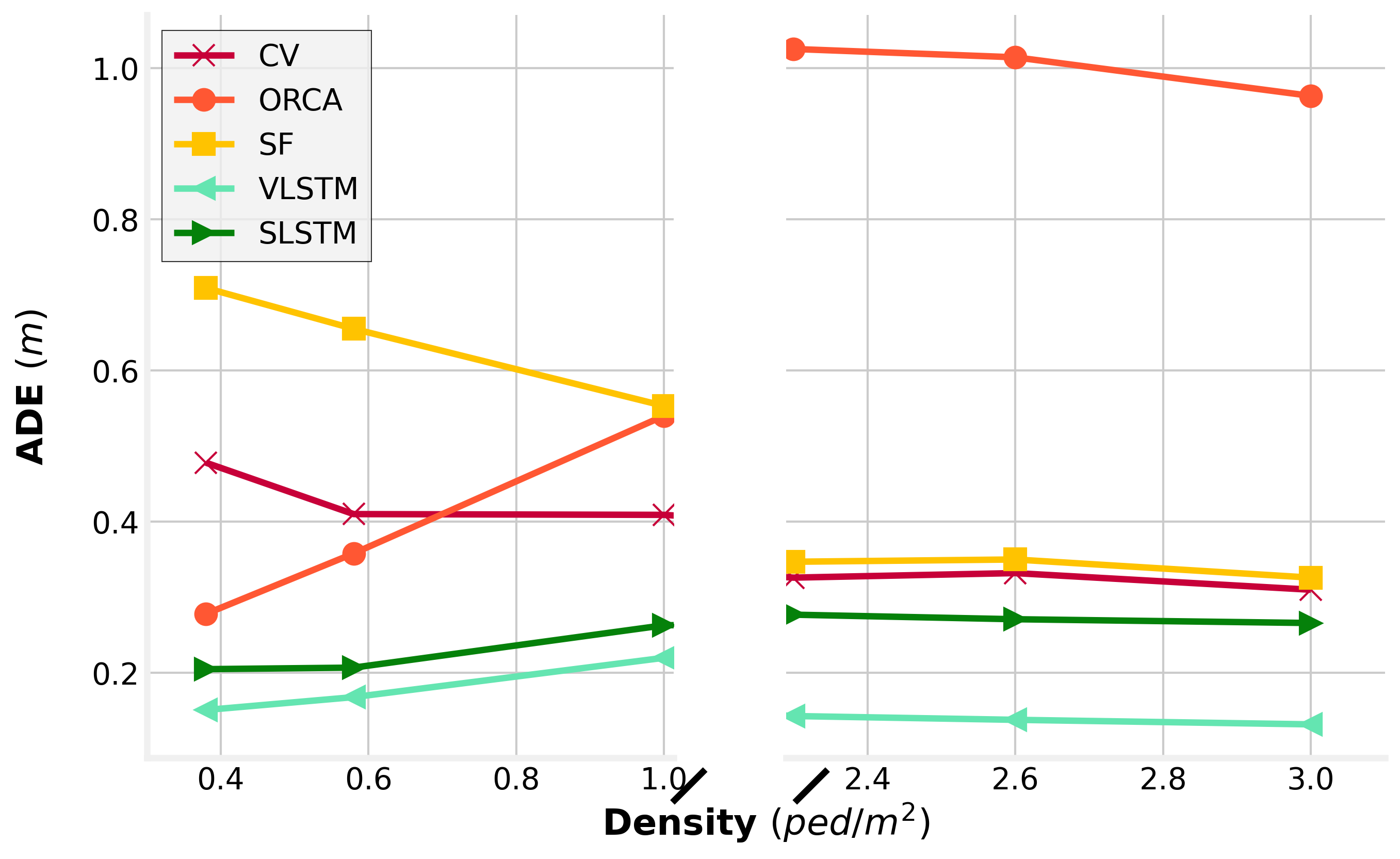}
        \caption{High-density data with $T_p=4.8$ sec.}
    \end{subfigure}
    \caption{Distance error-metric (ADE) for high-density datasets.}
    \label{ADE_high}
\end{figure}

In the next step, the same analysis of the different approaches with different prediction horizons is done for the high-density data. The results are presented in Fig.~\ref{ADE_high}.

As in Fig.~\ref{ADE_low}, the algorithms consistently outperform the models, irrespective of the dataset density. However, it is noteworthy that, surprisingly, the VLSTM exhibits superior performance compared to the more complex SLSTM. Additionally, it is notable that the error for SLSTM tends to increase with rising density, whereas for the VLSTM, it is lower within the higher density range of 2-3 ped/m$^2$. This unexpected outcome suggests that higher complexity does not necessarily lead to improved results when dealing with high densities. This assumption gains further support from the observed progression of the ADE for the CV approach, where the error appears to decrease with higher densities.

Similar to the results presented in Fig.~\ref{ADE_low}, we observed that the SF model performs better for longer prediction horizons compared to other approaches. Conversely, the ORCA model struggles to make accurate predictions at densities exceeding 2 ped/m$^2$. This limitation is attributed to the "freezing problem" highlighted in Luo et al. \cite{luo2018porca}, which becomes prominent at higher densities. For the longer prediction time, the performance of ORCA declines with higher density, while the performance of the SF improves with increasing density.

\subsection{Collision metrics}
\label{collision}
In this section, we will address an important challenge regarding collisions. In certain instances, pedestrians fail to avoid one another, contrary to what one would expect. This leads to overlapping or collision events, contravening one of the most critical physical criteria that a realistic prediction should satisfy. To assess the magnitude of this phenomenon, we will employ the two collision metrics Col and ITTC described in Section 3.

\subsubsection{Distance-based collision metric}
\label{Distance-based collision metric}
An essential aspect of the collision metrics is accurately defining when unrealistic behaviors, such as overlapping, occur. Hence, the shape of the pedestrians plays a crucial role. In the following, we present predictions made for two distinct radii: 0.1 meters and 0.2 meters.
The y-axis represents the percentage of predictions demonstrating collisions, as determined by the distance-based collision metric. The black line has not been presented in the figures before and it displays the percentage of collisions in the real datasets.

\begin{figure}[htbp]
    \centering
    \begin{subfigure}[t]{0.4\textwidth}
        \centering
        \includegraphics[width=2.3in]{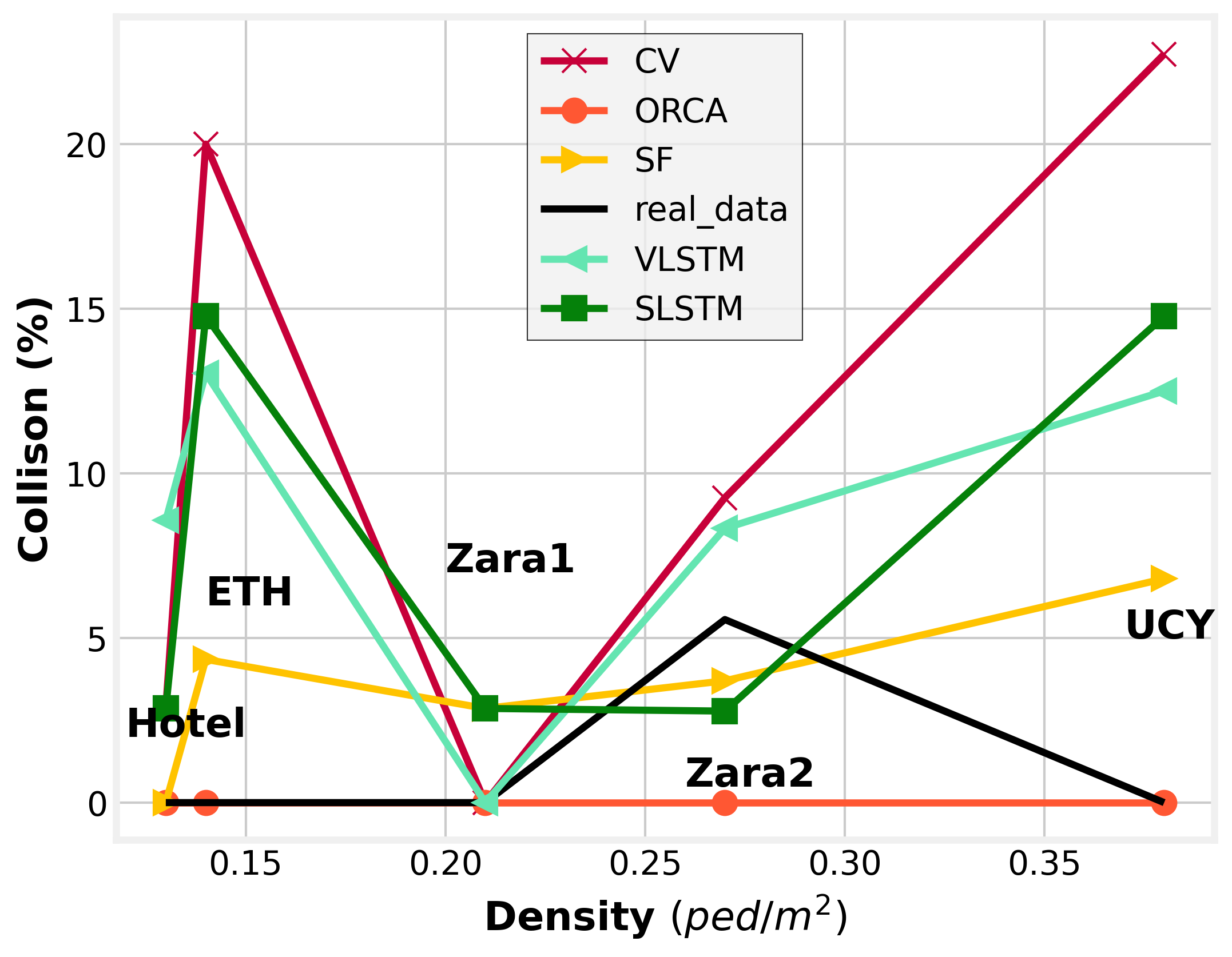}
        \caption[1]{Low-density data, $R=0.1$.}
    \end{subfigure}%
    \hspace{1cm}
    \begin{subfigure}[t]{0.4\textwidth}
        \centering
        \includegraphics[width=2.3in]{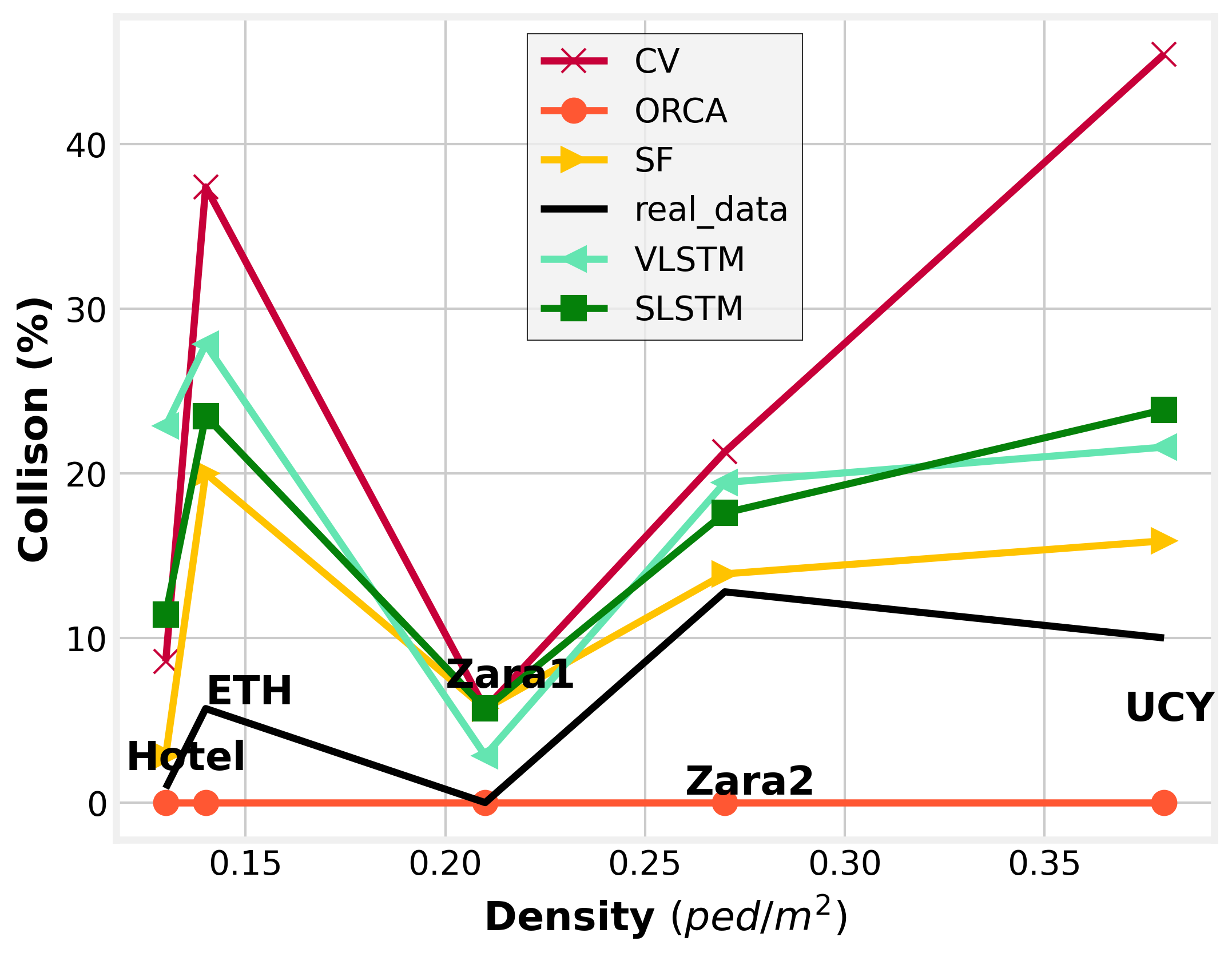}
        \caption{Low-density data, $R=0.2$.}
    \end{subfigure}
    \caption{Distance-based collision metric for different radius sizes.}
    \label{col_high_r1}
\end{figure}

In Fig.~\ref{col_high_r1}, it is evident that, as anticipated, the percentage of collisions is significantly higher for the larger radius. This observation holds true for all approaches and the real data, with the exception of the ORCA model, which demonstrates no collisions for either radius. Regarding the real data, collisions at a radius of 0.1 meters only occur within the Zara2 dataset, whereas collisions occur in the HOTEL, ETH, Zara2, and UCY datasets for a radius of 0.2 meters. This finding is surprising because collisions should not occur in the real trajectories. Collision between pedestrian occur very rarely in the real world. However, upon animating the real trajectories and plotting the colliding trajectories, it becomes apparent that the distance-based collision metric has a drawback: it sometimes considers grouping behavior as collision behavior. When pedestrians walk in groups, they occasionally come so close together that even at a radius of 0.1 meters, they sometimes overlap. In the next Fig.~\ref{col_high} the distance-based collision metric for the high-density data is presented.

\begin{figure}[htbp]
    \centering
    \begin{subfigure}[t]{0.4\textwidth}
        \centering
        \includegraphics[width=2.3in]{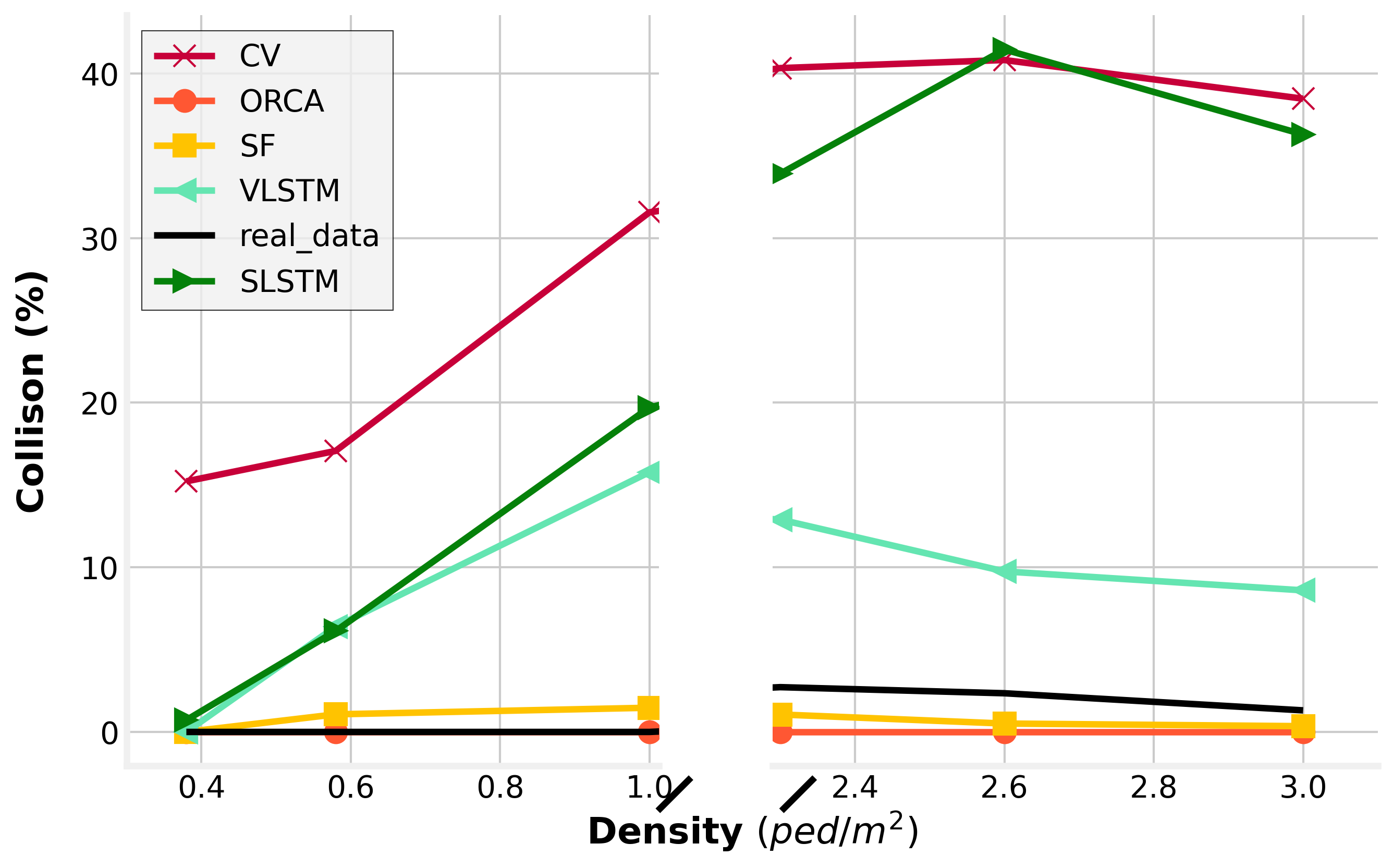}
        \caption[1]{Bidirectional data, $R=0.1$.}
    \end{subfigure}%
    \hspace{1cm}
    \begin{subfigure}[t]{0.4\textwidth}
        \centering
        \includegraphics[width=2.3in]{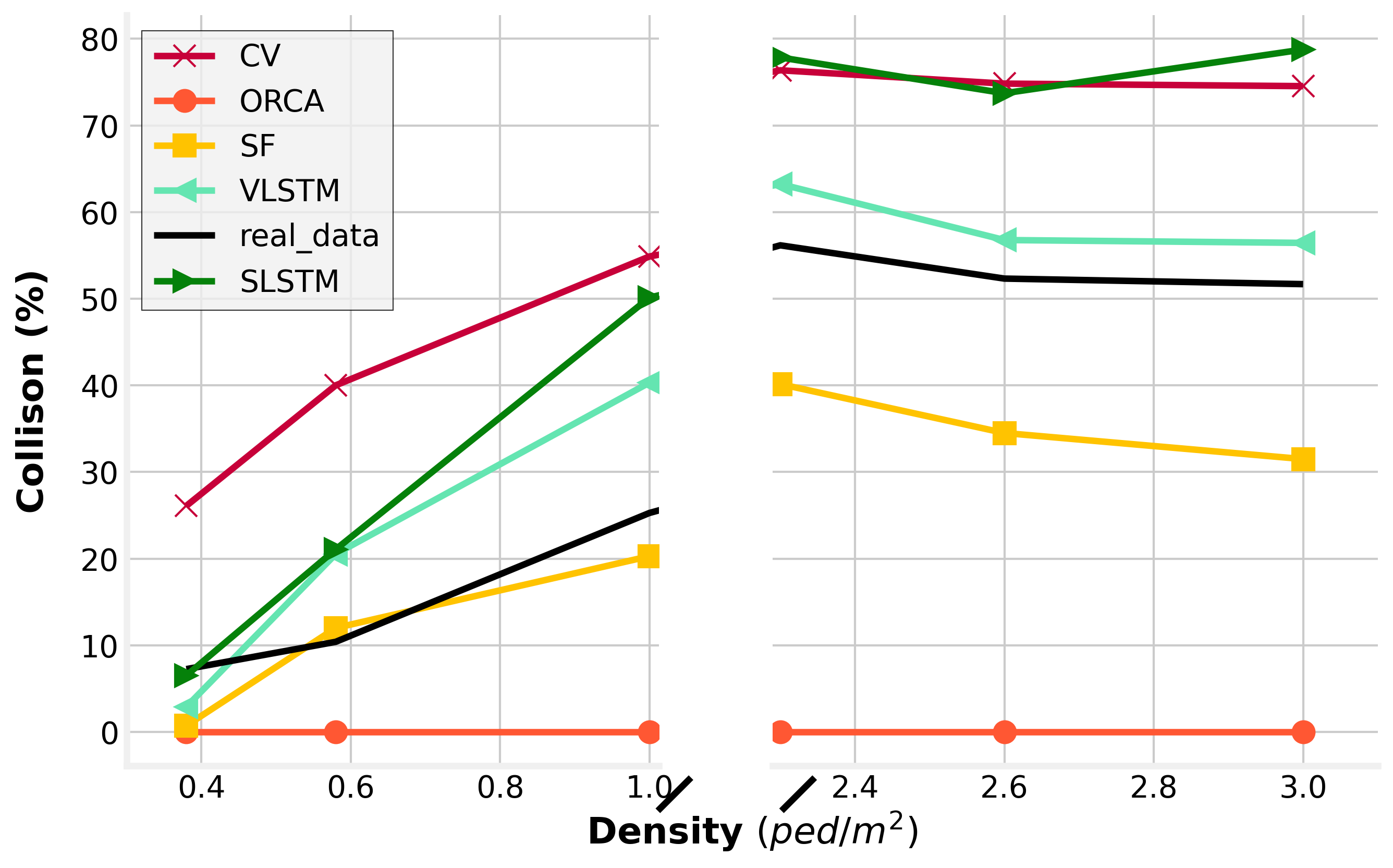}
        \caption{Bidirectional data, $R=0.2$.}
    \end{subfigure}
    \caption{Distance-based collision metric for different radius sizes.}
    \label{col_high}
\end{figure}

Substantial disparities are evident between the two diagrams representing collision occurrences at varying radii. The predicted collision frequency is nearly double at a radius of 0.2 meters as compared to a smaller radius. The black line, indicative of actual collision data, suggests a negligible number of collisions at high-density datasets for a radius of 0.1 meters. However, when the radius is increased to 0.2 meters, the predictions indicate almost 50 \% more collisions in higher densities. These observations suggest that a radius of 0.2 meters is excessively large for such scenarios. Pedestrians at these densities are in such close proximity that they often overlap when represented as circular objects. The ORCA model predicts no collisions for either radius, but this results in extraordinarily high ADE values for large prediction horizons, as depicted in Fig.\ref{ADE_high}, right panel. The CV model underperforms and displays the highest collision rate in its predictions. The SLSTM shows comparable performance at lower densities as demonstrated in Fig.~\ref{col_high}, but its effectiveness diminishes at higher densities. The VLSTM model exhibits superior performance to the SLSTM in terms of collision metrics.
The percentage of predicted collision of the SF model most closely aligns with actual trajectories.

\subsubsection{TTC-based collision metric}
\label{TTC-based collision metric}
In this section, we will discuss the results of the predictions centered around the ITTC. As observed in the preceding figures, the x-axis represents the density of the data. On the other hand, the y-axis denotes ITTC of the predictions (see equation~\ref{col-I}). It is important to note that the maximum TTC value was arbitrarily set to 12 seconds. A higher TTC can be interpreted as an indication of a prediction that is not at risk of collision. Consequently, a lower ITTC value is preferable for collision avoidance compared to a higher one.

\begin{figure}[htbp]
    \centering
    \begin{subfigure}[t]{0.4\textwidth}
        \centering
        \includegraphics[width=2.3in]{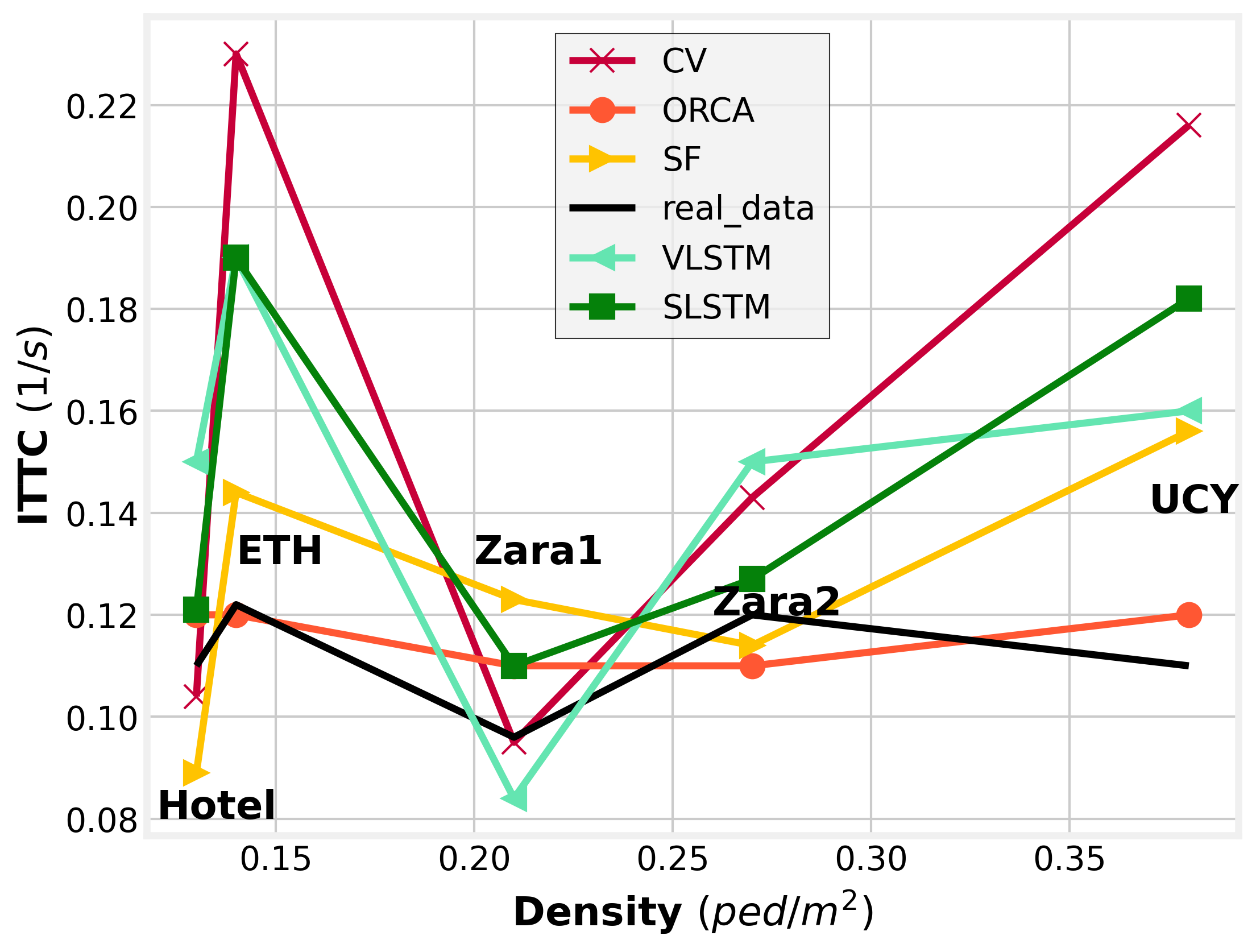}
        \caption[1]{Low-density data, $R=0.1$.}
    \end{subfigure}%
    \hspace{1cm}
    \begin{subfigure}[t]{0.4\textwidth}
        \centering
        \includegraphics[width=2.3in]{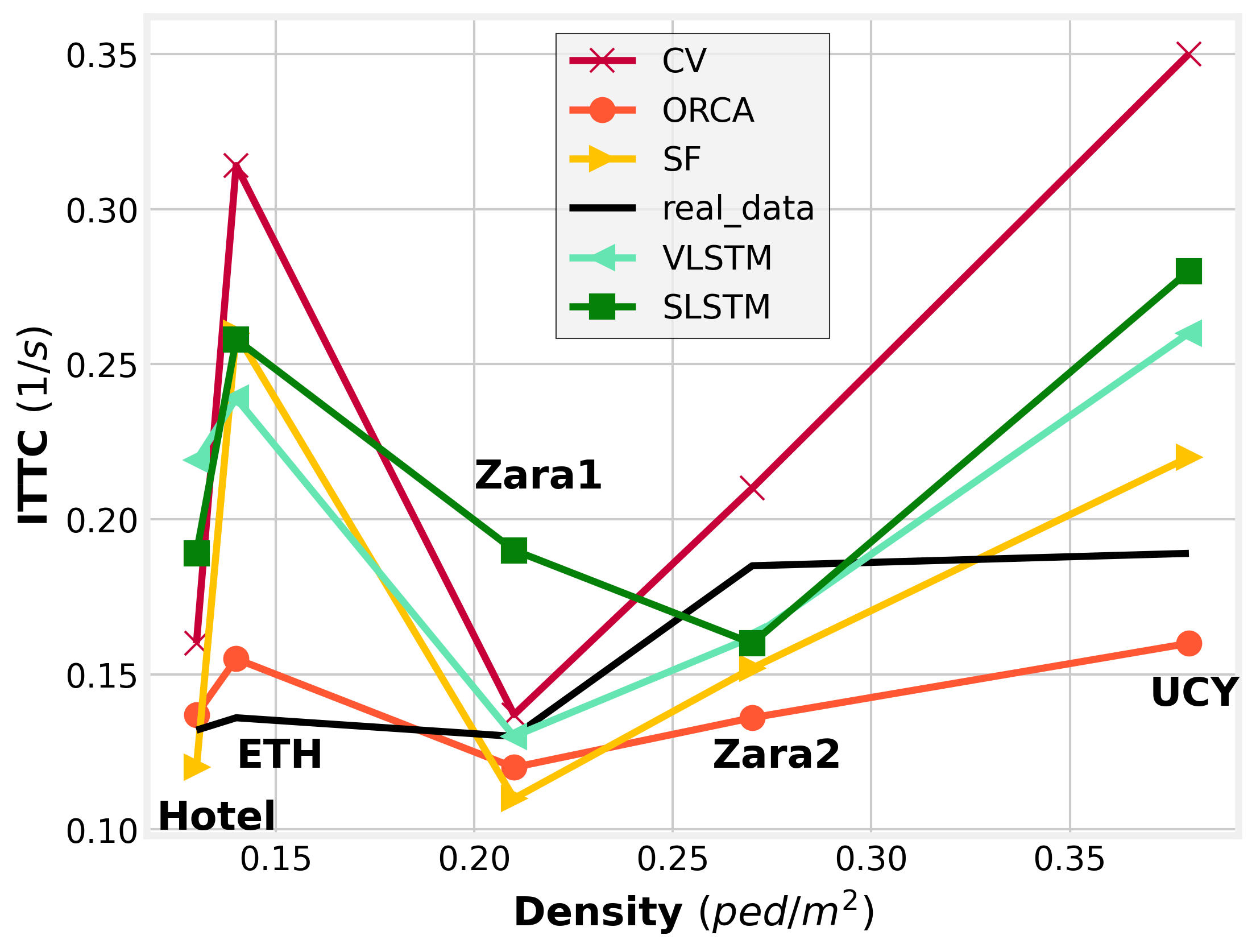}
        \caption{Low-density data, $R=0.2$.}
    \end{subfigure}
    \caption{TTC collision metric for different radius sizes.}
    \label{ttc_high_r1}
\end{figure}

Upon initial observation of Fig.~\ref{ttc_high_r1}, it is evident that the differences between Fig.~\ref{ttc_high_r1} (a) and (b) are considerably smaller compared to those seen in the distance-based collision metric. This observation suggests that the TTC-based collision metric is less sensitive to changes in the radius.

\begin{figure}[htbp]
    \centering
    \begin{subfigure}[t]{0.4\textwidth}
        \centering
        \includegraphics[width=2.3in]{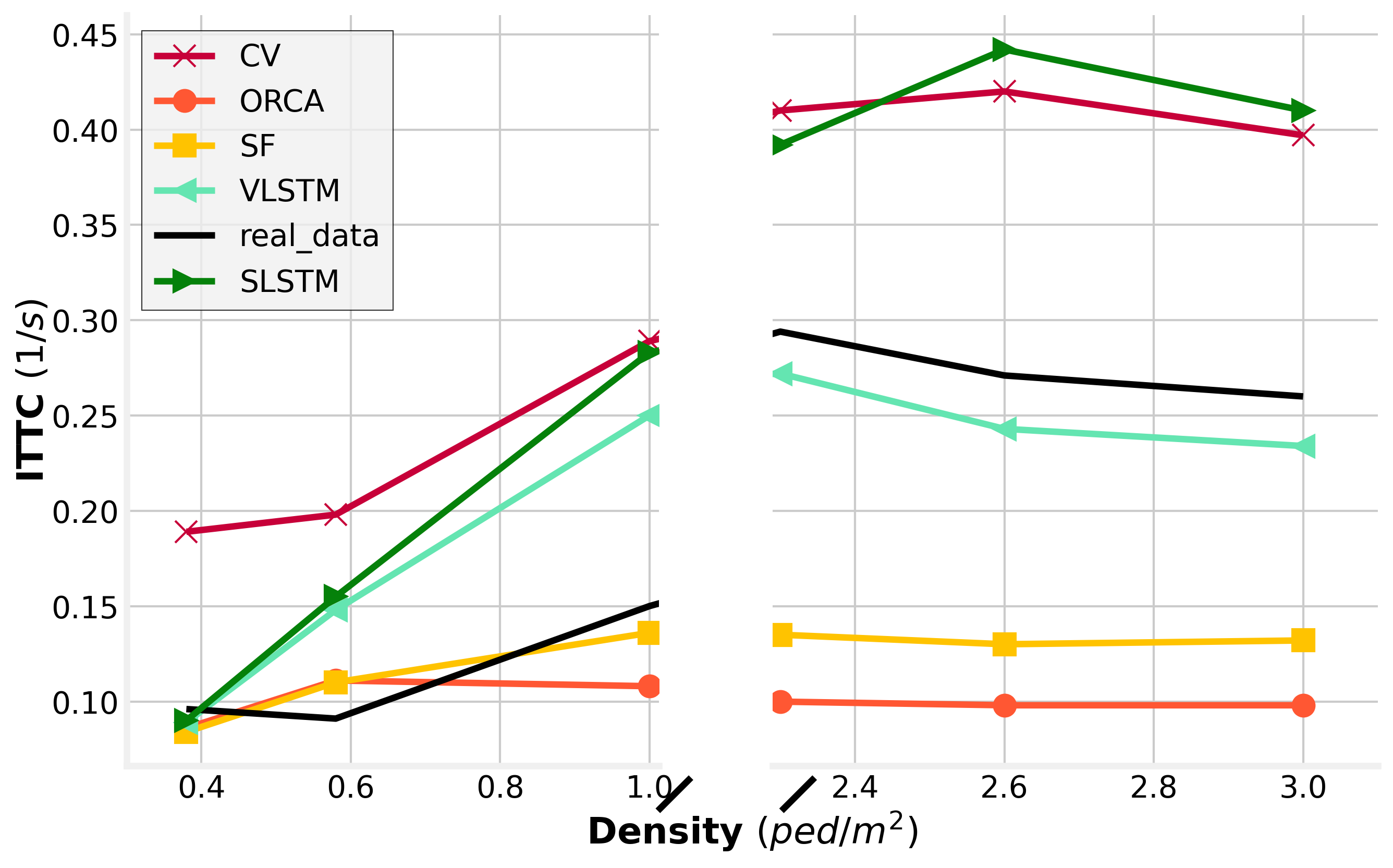}
        \caption[1]{High-density data, $R=0.1$.}
    \end{subfigure}%
    \hspace{1cm}
    \begin{subfigure}[t]{0.4\textwidth}
        \centering
        \includegraphics[width=2.3in]{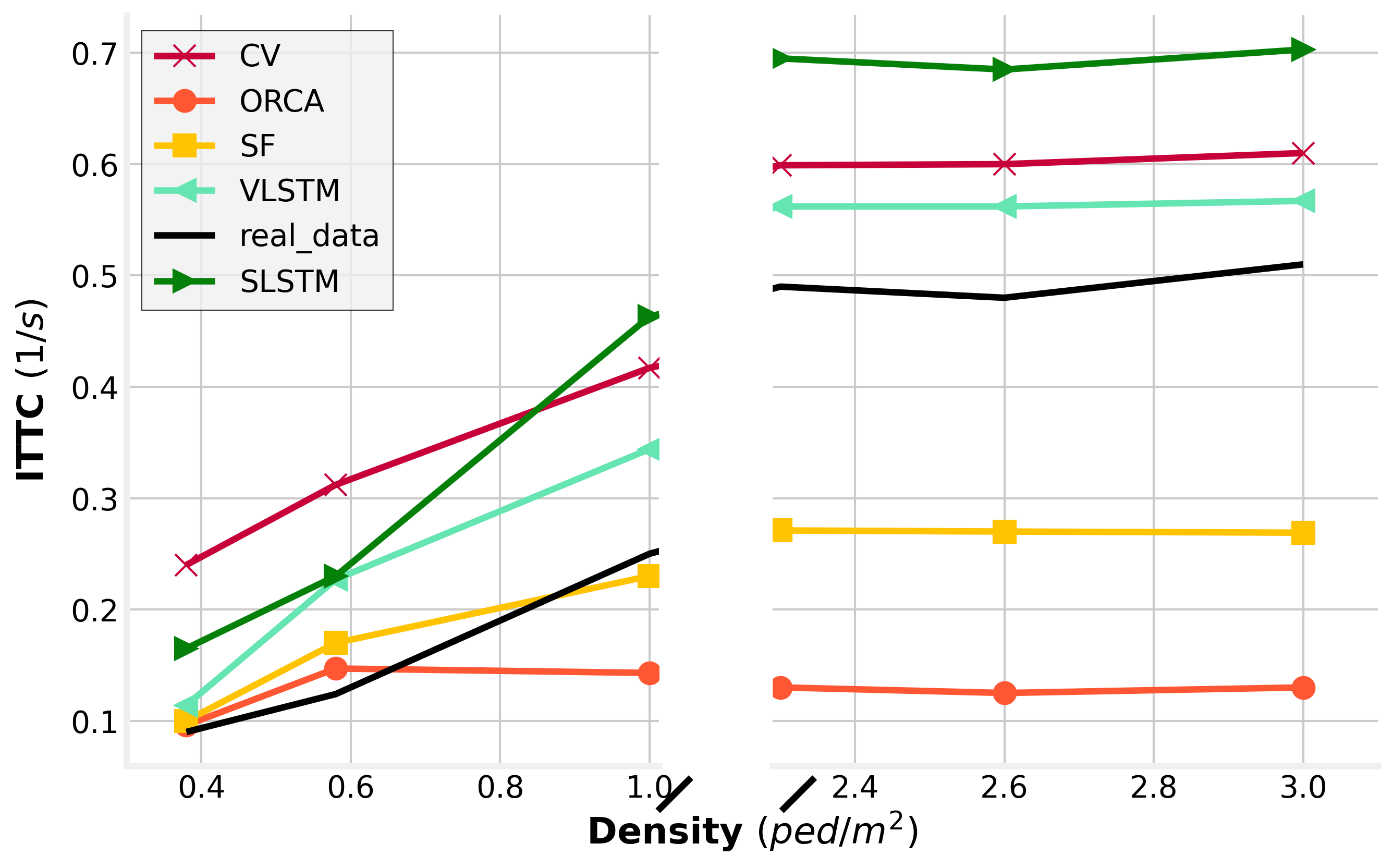}
        \caption{High-density data, $R=0.2$.}
    \end{subfigure}
    \caption{TTC collision metric different radius sizes.}
    \label{ttc_high}
\end{figure}

Despite this difference, the overall trend of the lines bears a resemblance to that seen in Fig.~\ref{col_high_r1}. 
However, the information provided by the continuous collision metric is richer and appears to be more accurate. The KB models exhibit lower inverse ATTC values than the algorithms, with the CV model demonstrating the least optimal performance. Notably, the inverse ATTC of ORCA's prediction most closely aligns with the actual trajectories. Proceeding further, we will present the ITTC for high-density data in Fig.~\ref{ttc_high}.

Again, it is noteworthy, that the variations between Fig.~\ref{ttc_high} (a) and (b) are not as substantial as those seen in Fig.~\ref{col_high}. In the latter virtually no collisions are observable in the actual trajectories (black line) for a radius of 0.1 meters, yet the collision rates for a radius of 0.2 meters are markedly high. In contrast, in Fig.~\ref{ttc_high}, the difference between (a) and (b) for real trajectories is much less pronounced. Despite this, the trends in Fig.~\ref{col_high} and Fig.~\ref{ttc_high} do not exhibit significant discrepancies, with the KB models consistently surpassing the algorithms and the SLSTM model underperforming in comparison to the VLSTM. In some instances, SLSTM even demonstrates worse performance than the CV model.

\subsection{TTC Metric for improving performance of the algorithm}
\label{TTC Metric for improving performance of the algorithm}
The preceding section demonstrates the promising attributes of the ITTC collision metric. It provides reasonable results that have the advantage of continuity and exhibit less dependence on the pedestrians' shape compared to the distance-based collision metric Col. As such, we have incorporated the TTC into the cost function to see if the prediction can be improved. The cost function quantifies the disparity between the prediction and the real observation. It provides a single indicator $L_i$ that will be minimized during the training. The idea is to penalize predictions with exceptionally low TTC values.
Keeping all configurations the same ($T_p=4.8$ seconds and R=0.2), we have integrated the TTC into the cost function of the SLSTM, as illustrated in equation~\ref{m4}

% \begin{equation}
%     \textbf{L}= \sum_{i}\|x_i - \hat{x}_i\|^2 + \lambda  f(\tau).
% \label{m4}
% \end{equation}

\begin{equation}
    \textbf{L}_i = \sum_{t=1}^T\|x_i(t) - \hat{x}_i(t)\|^2 + \lambda \sum_{t=1}^T f(\min_{j \ne i} \{\tau_{ij} \} ).
\label{m4}
\end{equation}

\begin{figure}[htbp]
    \centering
    \begin{subfigure}[t]{0.45\textwidth}
        \centering
        \includegraphics[width=2.5in]{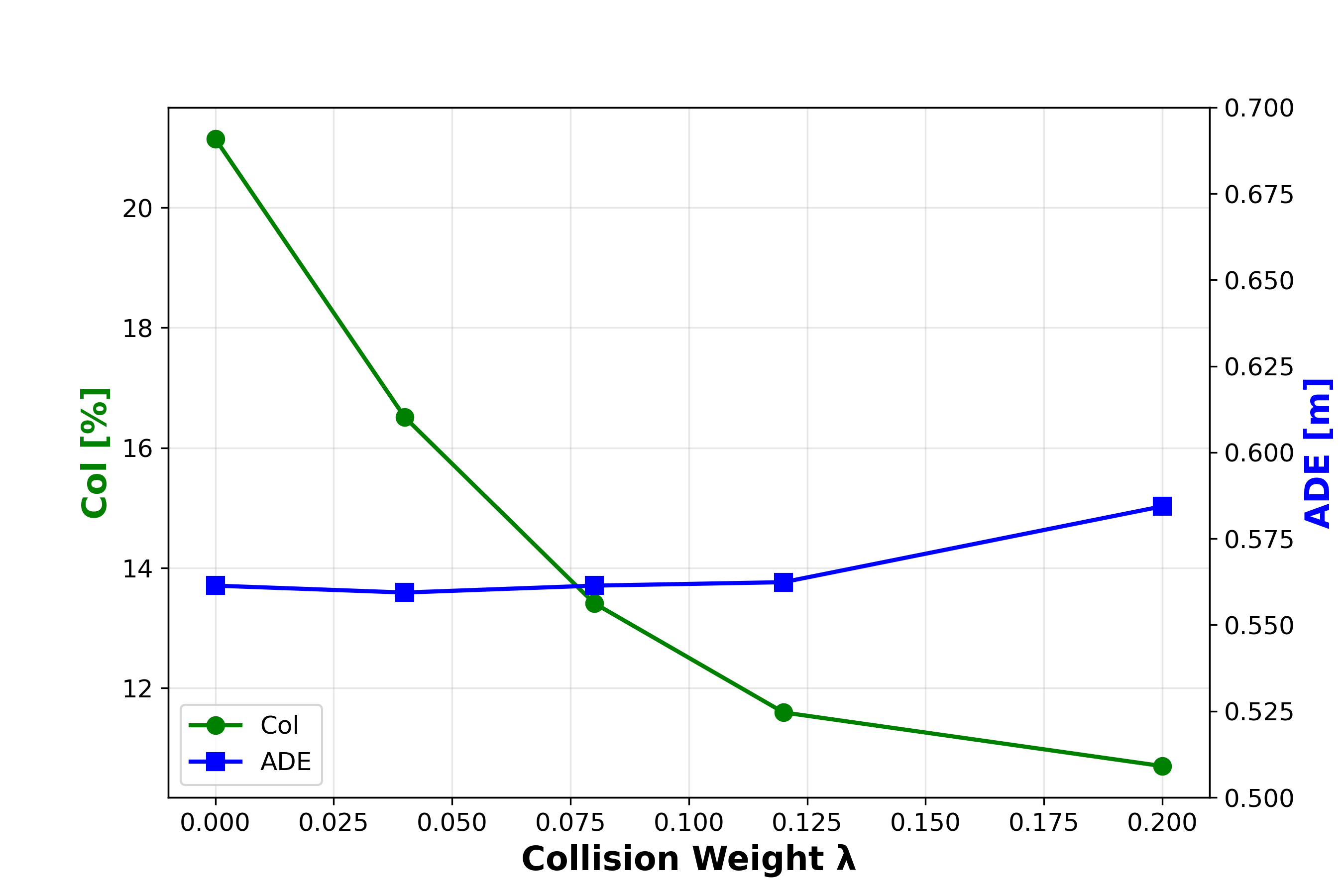}
        \caption[1]{Low-density data.}
    \end{subfigure}%
    \hspace{1cm}
    \begin{subfigure}[t]{0.45\textwidth}
        \centering
        \includegraphics[width=2.5in]{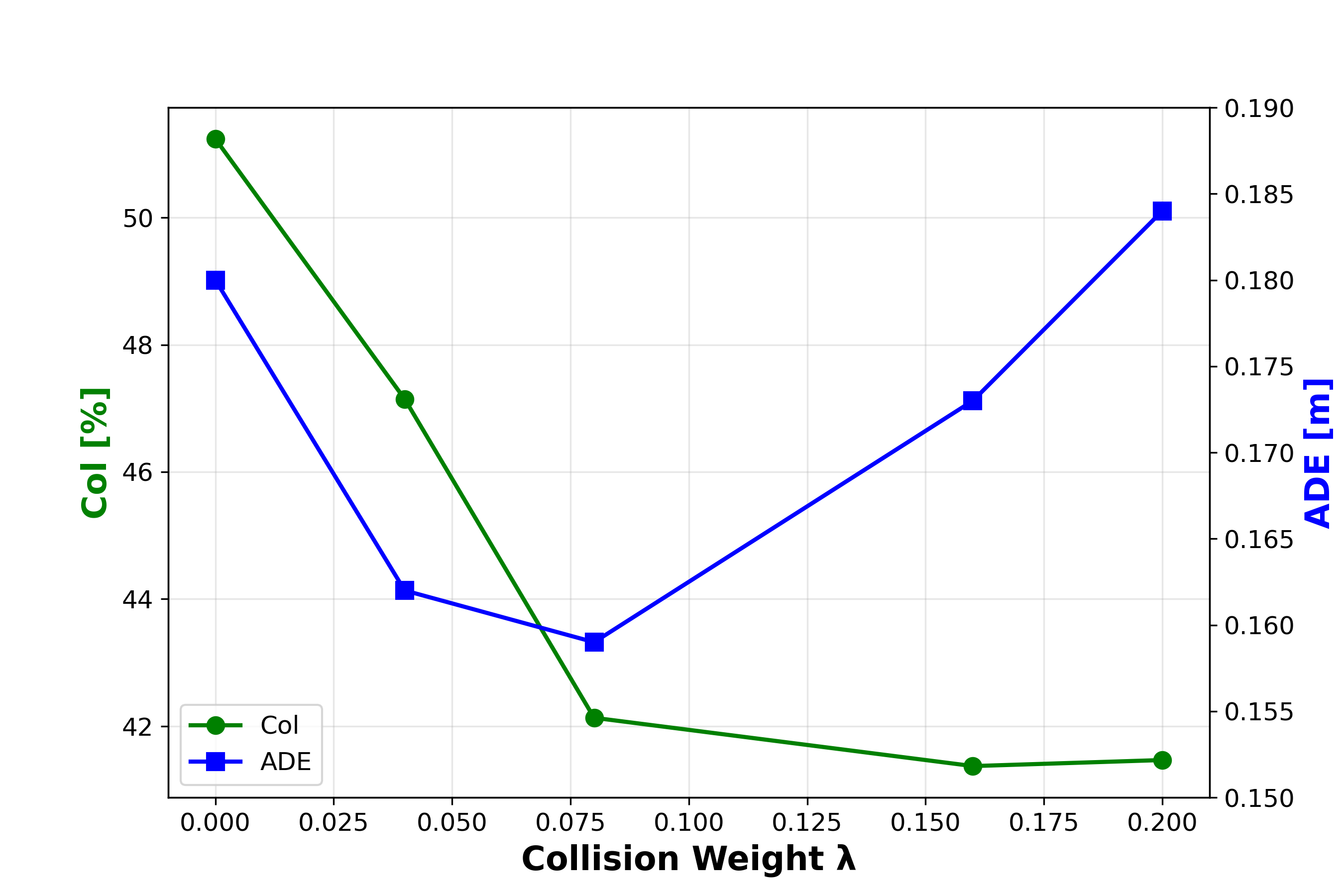}
        \caption{High-density data.}
    \end{subfigure}
    \caption{Illustrating the improvements achieved by incorporating TTC into the training function for both low-and high-density data.}
    \label{improving}
\end{figure}

The first part of the equation~\ref{m4} shows the distance-based metric ADE that is used for training. It compares the actual position of the pedestrian $x_i$ to the predicted one $\hat{x_i}$. The subsequent segment utilizes a sigmoid penalty function $f$ (see Equation~\ref{f_ttc}), which results in a high penalty for low TTC values
\begin{equation}
    f(\tau) = \frac{1}{1 + e^{s (\tau - \delta)}} ,
\label{f_ttc}
\end{equation}
where $s$ and $\delta$ are slope and threshold parameters, respectively. \\
The parameter $\lambda\ge0$ determines the weight to be given to the second part of the cost function. In Fig.~\ref{improving} the results of the predictions in terms of ADE and Col are shown for different settings of $\lambda$.

\begin{figure}
  \centering
  \vspace{-3cm}
  \begin{minipage}[b]{0.49\textwidth}
    \includegraphics[width=\textwidth]{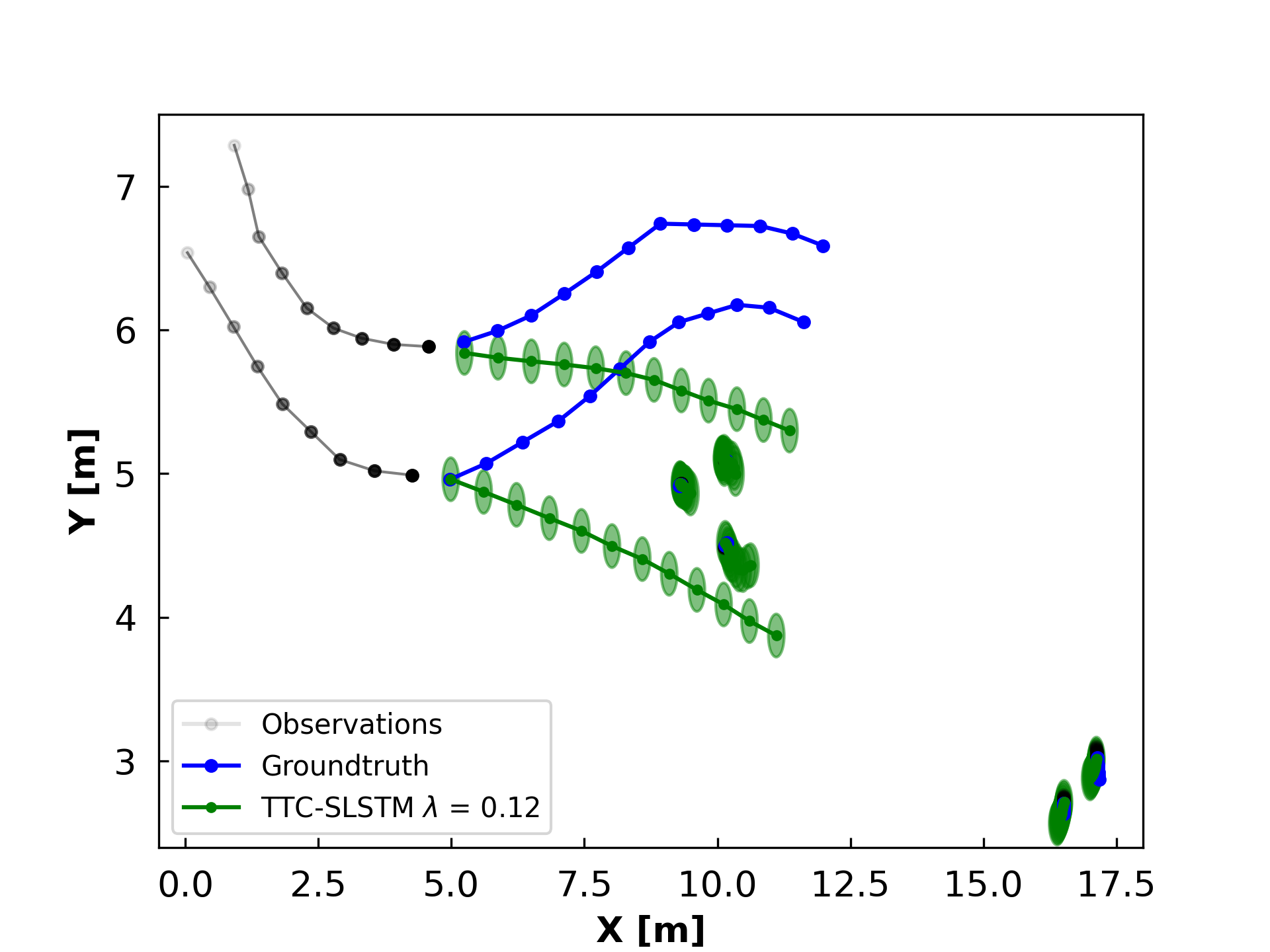}
    
  \end{minipage}
  %\hfill
  \begin{minipage}[b]{0.49\textwidth}
    \includegraphics[width=\textwidth]{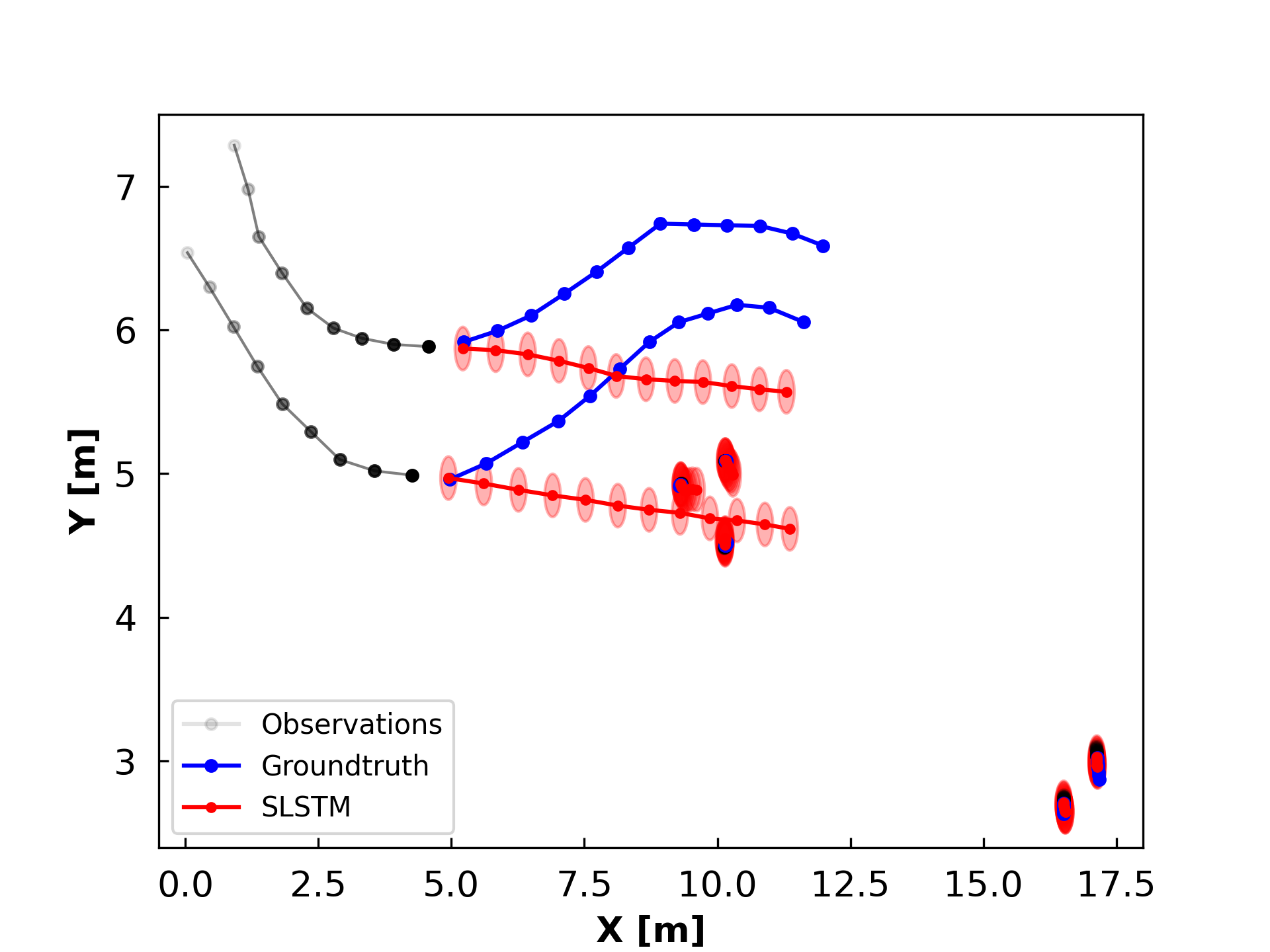}
    
  \end{minipage}
  \vspace{0.cm} % You can adjust the space between the rows as needed

  \begin{minipage}[b]{0.49\textwidth}
    \includegraphics[width=\textwidth]{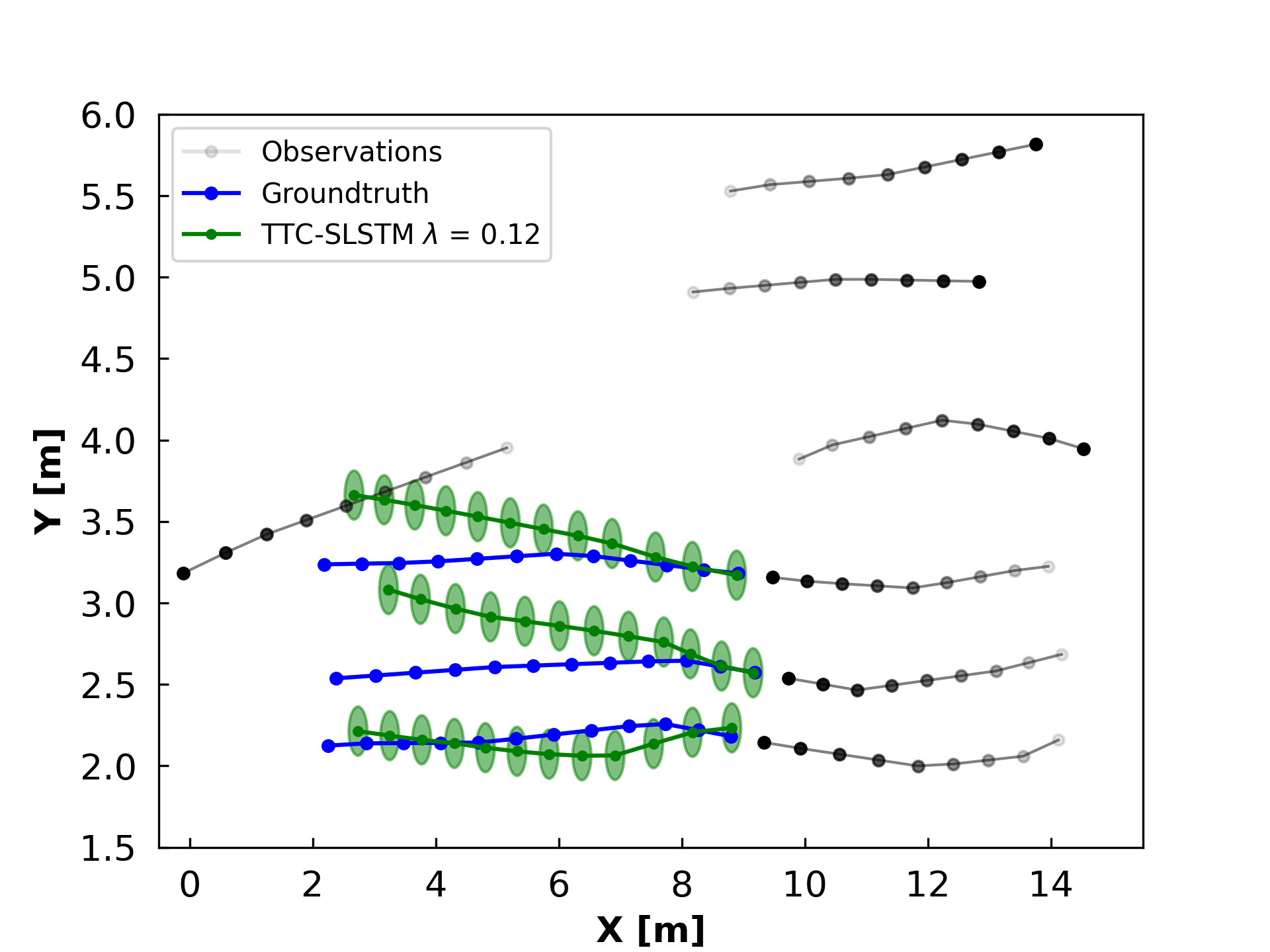}
    \subcaption{Prediction with TTC in cost function.}\label{fig:img3}
  \end{minipage}
  %\hfill
  \begin{minipage}[b]{0.49\textwidth}
    \includegraphics[width=\textwidth]{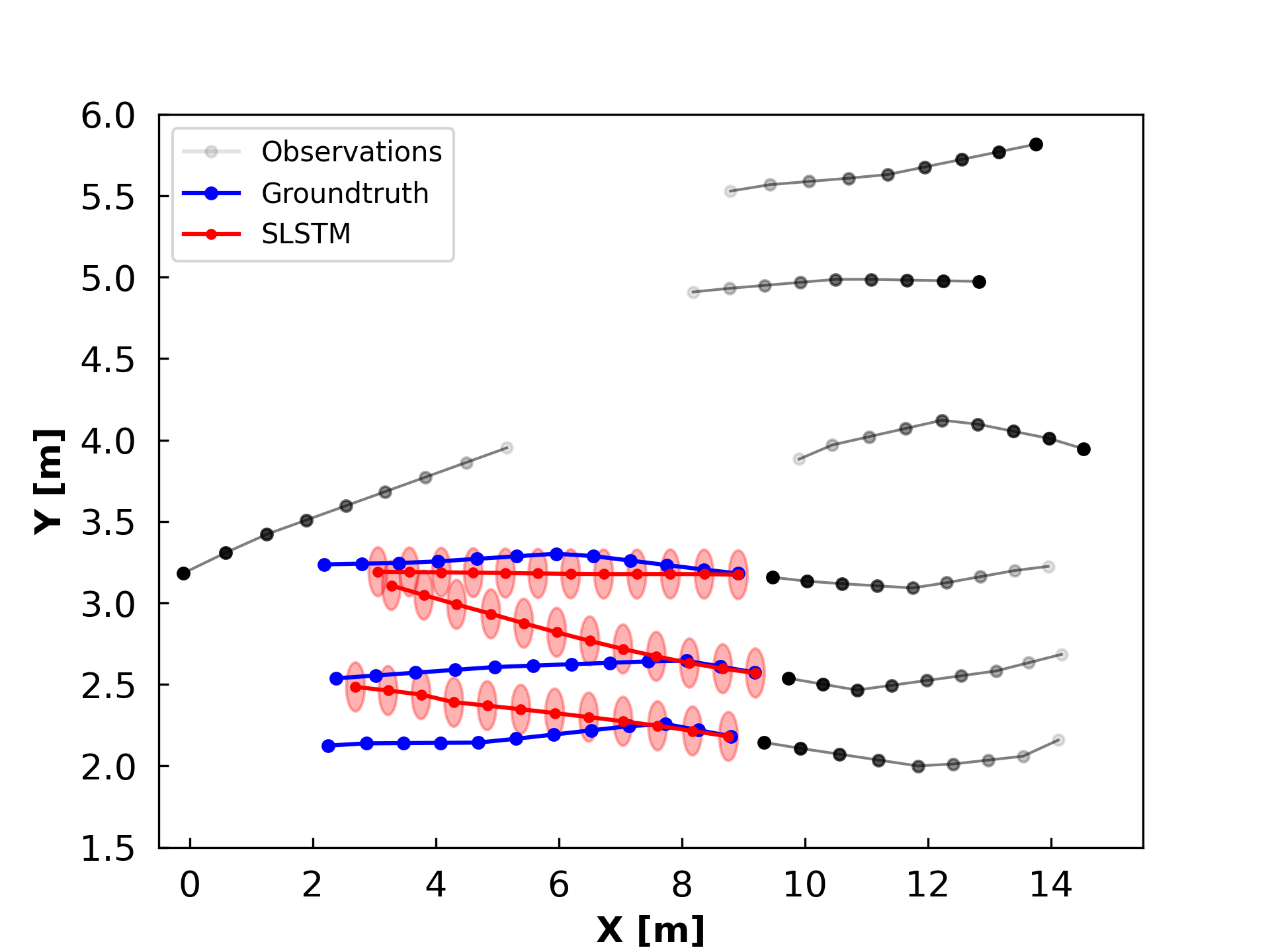}
    \subcaption{Prediction with SLSTM.}\label{fig:img4}
  \end{minipage}

  \caption{Example of trajectory predictions based on the algorithm that was trained with TTC in the cost function and the SLSTM.}
  \label{predictions}
\end{figure}

The green line represents the Col metric, while the blue line shows ADE. In the first observation, the value of $\lambda$ is zero, which means that this model is identical to the SLSTM without a TTC term in the cost function. It can be clearly shown, that the TTC term in the cost function helps to improve avoidance behavior. There is a strong relationship between the value of $\lambda$ and the number of collisions in the predictions. In the case of low-density data, it is possible to halve the number of collisions without an accompanying increase in ADE. For the high-density data, a reduction in the Col metric by 20 \% is achievable, which also leads to a decrease in ADE. The enhancements afforded by the incorporation of TTC into our algorithm are discernible both quantitatively and qualitatively. When visualizing the predicted trajectories, those produced by the TTC-incorporated model appear more realistic, exhibiting superior collision avoidance characteristics. Fig.~\ref{predictions} presents two scenes from the low density data. On the left side the predictions made by the algorithm trained with TTC are shown and on the right side the predictions with  the SLSTM.

In both scenes, the SLSTM's predictions result in collisions. In the upper image, pedestrians do not navigate around the stationary individuals in the middle. Meanwhile, in the lower image, the pedestrians form a group walking so close to each other that it registers as a collision. In the predictions generated using TTC, pedestrians within a group maintain a greater distance from each other. Additionally, in the upper image, pedestrians navigate successfully around those stationary in the center, further demonstrating the benefits of incorporating TTC into trajectory predictions.

\section{Conclusion}
In this paper, we have conducted a detailed empirical analysis comparing various pedestrian trajectory prediction approaches. The investigation underscores that the task of predicting pedestrian trajectories is intrinsically complex, with different densities posing additional challenges. While the SLSTM demonstrates excellent performance in low-density scenarios, it struggles to maintain similar accuracy in high-density situations. A particular limitation of the DB algorithms, namely, a significant incidence of collisions in high-density predictions, is addressed by introducing an innovative continuous collision metric that calculates the time-to-collision between pedestrians. This new metric presents a valuable instrument to assess the performance of the approaches, enhancing the overall trajectory prediction accuracy realism feature in terms of hardcore body exclusion. Future work will continue to optimize this metric and further explore its potential in improving the safety and efficiency of pedestrian movement predictions. This investigation thus serves as a foundation for refining and expanding the capabilities of data-based algorithms for predicting and modelling pedestrian behavior at high densities.

\section*{Acknowledgment}
  The authors acknowledge the Franco-German research project MADRAS funded in France by the Agence Nationale de la Recherche (ANR, French National Research Agency), grant number ANR-20-CE92-0033, and in Germany by the Deutsche Forschungsgemeinschaft (DFG, German Research Foundation), grant number 446168800. 
\section*{Author Contributions}
R.K and A.T formulated the conceptualization and methodology. R.K wrote the manuscript, performed the formal analysis, and visualized the results. R.K and T.D developed the software and performed the data curation. A.T assumed the supervision and funding acquisition. All authors reviewed the article. 


%%       \citet{<label>} ==> Jones et al. [21]
%%       \citep{<label>} ==> [21]
%%

\bibliographystyle{elsarticle-num-names} 
\bibliography{biblio.bib}


%\begin{thebibliography}{00}



%\end{thebibliography}
\end{document}